\documentclass[journal]{IEEEtran}
\usepackage{cite}
\ifCLASSINFOpdf
  \usepackage[pdftex]{graphicx}
  \graphicspath{{../pdf/}{../jpeg/}}
  \DeclareGraphicsExtensions{.pdf,.jpeg,.png}
\else
  \usepackage[dvips]{graphicx}
  \graphicspath{{../eps/}}
  \DeclareGraphicsExtensions{.eps}
\fi
\usepackage{amsmath}
\usepackage{amsfonts}
\usepackage[dvipsnames]{xcolor}
\usepackage{amssymb}
\usepackage{array}
\usepackage{multirow}
\usepackage{longtable}
\usepackage[english]{babel}
\usepackage{amsthm}
\usepackage{scalerel}
\usepackage{hyperref}
\newtheorem{remark}{Remark}
\usepackage{textcomp}
\usepackage{url}
\usepackage[
linesnumbered,ruled,vlined]{algorithm2e}
\usepackage {algpseudocode}
\usepackage{algorithmicx}
\usepackage{algcompatible}

\makeatletter
\newcommand{\thickhline}{%
    \noalign {\ifnum 0=`}\fi \hrule height 1pt
    \futurelet \reserved@a \@xhline
}
\newcolumntype{"}{@{\hskip\tabcolsep\vrule width 1pt\hskip\tabcolsep}}
\makeatother
\ifCLASSOPTIONcompsoc
  \usepackage[caption=false,font=normalsize,labelfont=sf,textfont=sf]{subfig}
\else
  \usepackage[caption=false,font=footnotesize]{subfig}
\fi
\ifCLASSOPTIONcaptionsoff
  \usepackage[nomarkers]{endfloat}
 \let\MYoriglatexcaption\caption
 \renewcommand{\caption}[2][\relax]{\MYoriglatexcaption[#2]{#2}}
\fi
\let\MYorigsubfloat\subfloat
\renewcommand{\subfloat}[2][\relax]{\MYorigsubfloat[]{#2}}
\begin{document}
\title{Wide-Area Power System Oscillations \\
from Large-Scale AI Workloads}

\author{Min-Seung~Ko,~\IEEEmembership{Member,~IEEE,} and
Hao~Zhu,~\IEEEmembership{Senior~Member,~IEEE}

\thanks{This work has been supported by NSF grants 2130706 and 2150571 (\textit{Corresponding Author: Hao Zhu.})}
\thanks{M.-S. Ko and H. Zhu are with the Chandra Family Department of ECE, The University of Texas at Austin, Austin, TX 78712, USA. (e-mail: kms4634500@utexas.edu, haozhu@utexas.edu)}}

\markboth{IEEE Transactions on Power Systems}{}
\maketitle
 
\begin{abstract}
This paper develops a new dynamic power profiling approach for modeling AI-centric datacenter loads and analyzing their impact on grid operations, particularly their potential to induce wide-area grid oscillations. We characterize the periodic stochastic power fluctuations inherent to large-scale AI workloads during both the training and fine-tuning stages, driven by the state-of-the-art graphics processing unit (GPU) computing architecture design. % and distributed mini-batch processing cycles. 
These sustained, large power fluctuations, unlike conventional load ramping, act as persistent forcing inputs capable of interacting with and amplifying local and inter-area oscillation modes. Using the WECC 179-bus system and the NPCC 140-bus system, we have numerically studied the amplitude and variability of oscillatory responses under different factors. These factors include system strength, penetration level, fluctuation frequency range, individual datacenter size, geographical deployment, fluctuation suppression level, and workload ratio. Simulation results show that, notably, narrower fluctuation bands, larger single-site capacities, or dispersed siting can intensify oscillations across multiple modes. Our models and numerical studies provide a quantitative basis for integrating AI-dominant electricity demand into grid oscillation studies and further support the development of new planning and operational measures to power the growth of AI/computing load demands.
\end{abstract}

\section{Introduction}\label{sec1}
\IEEEPARstart{T}{he} rapid growth of large electrical loads, such as datacenters, cryptomining facilities, and transportation electrification, is fundamentally shifting the composition and characteristics of electricity demand. In particular, datacenters have emerged as a primary driver of large-load growth, due to the explosion of AI applications and large language models (LLMs) \cite{li2024unseen, quint2025practical}. By the end of 2030, datacenter electricity demand is projected to account for up to 12\% of total US electricity consumption \cite{arman20242024, aljbour2024powering}. Unlike conventional industrial loads, datacenter power profiles typically exhibit fast, sustained fluctuations with multi‑timescale variability driven by workload orchestration and scheduling changes \cite{li2025ai}. These unique patterns significantly challenge the grid operational paradigm in power balancing and dynamic control, and introduce new grid reliability considerations. Hence, there is an urgent need to develop a comprehensive framework for modeling datacenter electricity demand and assessing its potential impact on grid dynamic performance \cite{li2024unseen, quint2025practical}.

%{\hao do we need to add one number above to show the high-level penetration?}
% interconnection process & standard (ERCOT) / Interim LL interconnection process / resource adequacy
% VRT ~ dynamic impact
% Solution - Super cap + STATCOM?
% CUSTOMER-INITIATED LOAD REDUCTION EVENT (Dominion Energy)
% UPS / Lancium
% Datacenter campus (1GW ~ 5GW) + 360 MW Gas Power Plant 1 GW Battery (+ 1GW Solar)
% --> Large orchestrated campus
% ERCOT may curtail large loads in energency condition prior to residential/commercial load shed
% Panhandle 5GW datacenter
% WEST TEXAS LARGE DATACENTERS ~  765kV Transmission 

Along these lines, the majority of current research and industry assessments focus on large ramping behaviors and transient stability analysis. Abrupt workload initiation, shutdown, and inter-site transfers result in stepwise power changes that can significantly affect grid stability, especially with very high datacenter penetration \cite{jimenez2025data}. This issue has been already observed in major US interconnections such as electric reliability council of Texas (ERCOT), western electricity coordinating council (WECC), and Pennsylvania-New Jersey-Maryland (PJM) interconnection, where datacenter-initiated load shifts have caused measurable frequency and voltage disturbances \cite{robert2025event, Matt2025unplanned}. For example, graphics processing unit (GPU)-level power dynamics from AI workloads have been analyzed in \cite{li2024unseen}, which points out the potential stress on local grids. A combined modeling approach that incorporates both batch load patterns and AI-induced fluctuating demand is presented in \cite{jimenez2025data}, showing that sudden demand variations can be mitigated by adjusting the speed of frequency changes. 
To the best of our knowledge, most existing work on datacenter variability has focused solely on the effects of ramp power changes on power system transient responses. 

Notably, the grid dynamics arising from the rapid increase in datacenter demand and growing variability go far beyond ramping transients. With the latest explosions in large generative AI (GenAI) models and LLMs, next-generation super datacenters are increasingly being deployed and connected to power systems \cite{zhang2025integration}. Their GW-level power capacity not only significantly strains resource adequacy but also critically affects grid operations, leading to a vast increase in datacenter-induced power fluctuations that have not been seen in power systems thus far \cite{houle2025balance}. 
More recent technical assessments \cite{nerc2025char, quint2025practical} have started to recognize this urgent concern and pointed out that GenAI-centric datacenters can introduce periodic, sustained, and structured power fluctuations at the second level. This is because these next-generation datacenters are equipped with mega-scale GPU clusters and advanced computing architectures. They can, on a fast timescale, perform synchronized mini-batch processing cycles, job-level scheduling, and repetitive compute-to-communication phases \cite{patel2024characterizing}. If not managed, these resulting power fluctuations can act as external forcing signals to the grid, potentially triggering severe wide-area oscillations in future power systems with very high datacenter penetration. Unfortunately, quantitative models and systematic grid-level analyses of these fluctuations are largely absent from the literature. In particular, the frequency-selective and persistent nature of these GenAI-induced power fluctuations, as a forcing mechanism, is fundamentally different from ramp-induced transients, calling for a new modeling and analysis framework for the second-level oscillations. To this end, a rigorous quantitative evaluation of these fluctuations is essential to improve future datacenter-dominated grid operations and to power the integration of large-scale AI workloads.

To address this critical gap, this paper puts forth a comprehensive modeling approach to characterize the power fluctuations of large-scale AI workloads. Our goal is to enable an effective assessment of their potential to induce wide-area oscillations as the deployment of next-generation datacenters increases. We represent AI workloads with stochastic, periodic power consumption profiles, by considering the compute- and communication-dominated phase division within each cycle. Cycle durations can vary around a dominant time period, and phase power levels include both baseline shifts and fine-scale random deviations, with parameters tailored for training and fine-tuning workloads, respectively.
More importantly, we leverage the proposed models to perform a large-scale, grid-level analysis to identify a few deciding factors that can collectively shape the severity and variability of grid oscillations. By studying the oscillatory potentials in GenAI-dominant datacenter workloads, our work highlights an urgent need to enhance grid operations and stability amid the explosive growth in AI-driven demand.
The main contributions of this work are summarized as follows:
\begin{itemize}
    \item Establishing AI datacenter workloads as continuous, structured forcing sources beyond the existing ramping-centric models, to allow the analysis of their potential to induce inter‑area and local oscillations.
    \item Developing a stochastic AI workload model that captures both periodic patterns and fine-scale randomness, to produce realistic, time‑varying load profiles for grid studies.
    \item Systematically assessing the grid oscillation effects induced by large-scale AI workloads, revealing the aggregated resonance phenomena in a realistic power system.
    \item Identifying critical deployment factors and operational conditions, ranging from the system inertia, datacenter penetration level, fluctuation frequency characteristics, size of datacenters, geographical dispersion, fluctuation suppression, to workload ratio, in order to inform future grid planning strategies for AI-dominant grids.
\end{itemize}

The remainder of this paper is organized as follows. Section~\ref{sec2} discusses the characteristics and composition of AI workloads in new-generation datacenters. Section~\ref{sec3} proposes a stochastic modeling approach to capture the periodic, variable power consumption patterns of AI workloads. Section~\ref{sec4} details our case studies on the WECC 179-bus and NPCC 140-bus systems to assess oscillatory intensity and damping variability under various conditions, and the paper is wrapped up in Section~\ref{sec5}.

\section{Characteristics of AI Workloads and Datacenter Electricity Demand}\label{sec2}

\begin{figure}[t]
    \centering
    \includegraphics[width=0.98\columnwidth]{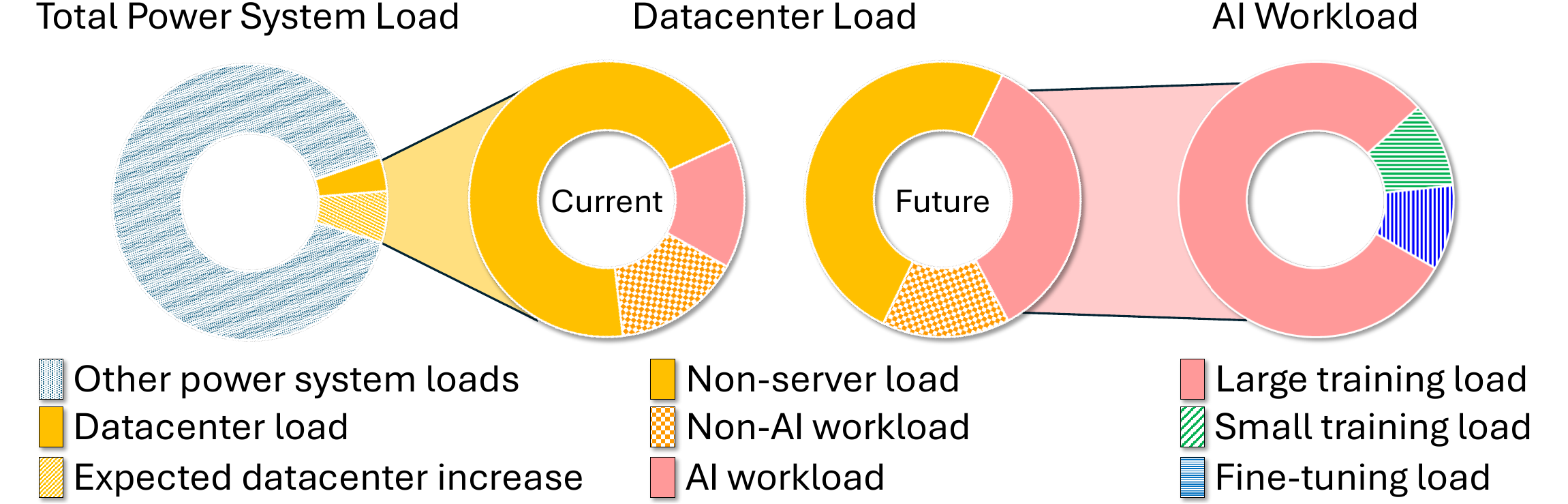}
    \caption{Example of a hierarchical AI workload composition.}
    \label{datcom}
\end{figure}

Large-scale and generative AI models, particularly LLMs, are rapidly emerging as dominant contributors to datacenter electricity demand, significantly reshaping both regional and interconnection-wide consumption patterns. Fig.~\ref{datcom} illustrates current and projected composition of datacenter electricity use and AI workloads. Recent studies \cite{davide2025energy, arman20242024, Matt2025unplanned} report that datacenters accounted for approximately 4\% of total US electricity demand in 2024, with this share expected to approach nearly 10\% by 2030. What is worse, certain North American regions have witnessed a concentrated datacenter development with potentially GW-level facilities, such as Texas, Virginia, and California. Consequently, the local penetration of datacenters in these regions will far exceed the average. Thus, there is an urgent need to improve the modeling of large-scale AI workload power profiles for grid-level studies.

Datacenter electricity consumption is primarily divided between non-computational infrastructure and server operations. Non-computational components, such as cooling, networking, and storage, typically account for 40--50\% of total consumption \cite{davide2025energy}. The remainder is consumed by server-side computational workloads. Conventional CPU-based server applications, such as web hosting, search indexing, and transactional processing, impose relatively small and stable demands.
In contrast, GPU-accelerated AI workloads incur substantially higher power consumption due to the large-scale parallelism. Current estimates indicate that AI workloads contribute roughly 10--20\% of total datacenter electricity use \cite{davide2025energy, buffard2025AI}. In AI-centric facilities, this fraction is expected to be considerably higher, potentially reaching 50\% in the future \cite{de2025artificial}.

From an operational standpoint, AI workloads are generally categorized into three stages: training, fine-tuning, and inference \cite{li2024unseen}. Training is the most compute-intensive stage, during which model parameters are optimized offline using comprehensive datasets for a pre-designed architecture. These jobs typically run continuously for days or weeks, resulting in sustained demands at peak loading. The fine-tuning stage follows training and adapts pre-trained models to task-specific applications using smaller datasets. Although fine-tuning has lower resource requirements and shorter compute time than training, particularly for GenAI models, it is often performed more frequently. As a result, it leads to lower but more sustained power consumption \cite{wang2023energy}.
Finally, the inference stage deploys fine-tuned models to perform real-time prediction tasks. Unlike the continuous, large-scale nature of training/fine-tuning, inference jobs are typically intermittent and event-driven in response to user queries, and their computational demands are much lower. 

Given the operational characteristics of AI workloads, the training and fine-tuning stages together constitute the majority of power consumption in AI-centric datacenters. Large-scale training tasks are typically executed as a single resource-intensive job distributed across hundreds or thousands of GPUs \cite{hu2021characterization}. Recent forecasts project that the peak power consumption of a single frontier AI training run could even reach 4-16 GW by 2030, a scale that would dominate the total electricity consumption of the hosted AI-centric facilities \cite{joshua2025scaling}. The remainder of active compute demand is primarily driven by smaller-scale training and fine-tuning tasks, which collectively sustain a high baseline load throughout operational cycles. This concentration is reinforced by an architectural split, with centralized GPU clusters for training and fine-tuning, and inference handled in lower-power zones or edge facilities \cite{avelar2023ai, brian2024inference}.

\section{Stochastic Power Profiling for AI Workloads}\label{sec3}

We develop a stochastic modeling framework to characterize the power consumption of training-centric AI datacenters. As established in Section~\ref{sec2}, inference workloads are intermittent, small‑scale, and hosted on separate low‑power infrastructure, limiting their impact on grid loading. Thus, the framework focuses on training and fine‑tuning, which together constitute the dominant and sustained component of demand. Both workloads are modeled within a unified stochastic framework while accounting for their distinct operational profiles. Although real measurement data for next-generation datacenters are not yet publicly available, our modeling choices are informed by the latest research on characterizing their electricity demand \cite{patel2024characterizing, li2024unseen, hu2024characterization, latif2024empirical, singhania2024methodology} and publicly available HPC power consumption datasets \cite{samsi2021supercloud, borghesi2023m100}.
%{\hao I think we should mention here that real data especially for next-generation super datacenters are not available, but we have followed latest research to develop these models.  }

\subsection{Training Workload Modeling}

\begin{figure}[t]
    \centering
    \begin{tabular}{cc}
    \includegraphics[width=0.45\columnwidth]{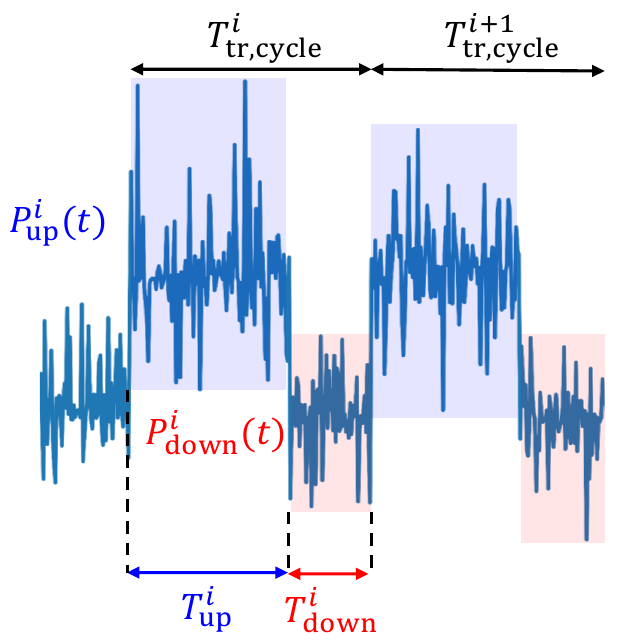} & \includegraphics[width=0.45\columnwidth]{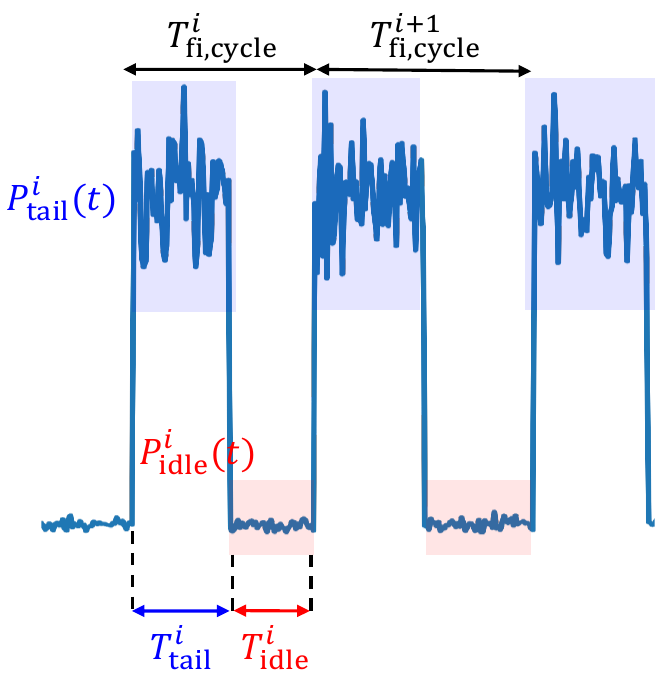} \\
    \small (a) &     \small (b)    
    \end{tabular}
    \caption{Example power consumption profiles for (a) training and (b) fine-tuning stages.}
    \label{siggen}
\end{figure}

AI training workloads in large-scale datacenters exhibit inherently periodic power consumption patterns due to the repetitive nature of mini-batch processing. Each iteration alternates between the compute-intensive, high-power \textit{up phases} and the low-power \textit{down phases}, driven by communication and synchronization needs. During the up phases, forward propagation, backward propagation, and gradient computation place heavy demands due to the inherent matrix/tensor operations and GPU kernel executions. In contrast, synchronization operations, memory transfers, and communication overhead result in lower power consumption during down phases \cite{patel2024characterizing, bridges2016understanding, hu2024characterization}. Beyond this periodicity, stochastic variations arise at two distinct time scales. Inter-iteration variability is driven by changes in cycle durations and phase power levels across iterations, resulting from GPU runtime jitter, operating-state changes, and kernel-scheduling delays \cite{sinha2022not, pham2020problems}. Intra-iteration variability arises within each phase itself, as various computational and communication tasks are sequentially executed, naturally inducing short-timescale deviations around the nominal power level \cite{hu2021characterization, hu2024characterization}.

%{\hao I suggested this before, but I still the following contains too much info. We didn't even get to define what the prob. model symbols mean... Could we please first use one sentence to define every symbol, and start another sentence to introduce and also JUSTIFY the probability models. Note that the two aspects are diff because the prob models are our assumption. We should try to differentiate them, so we don't run into the danger of upsetting the reviewrs. We have to do this for every part including the fine tuning... in addition, the subscript is typically used for index, and the superscript is more reserved for special notes. could we switch it everywhere? }
To precisely characterize this behavior, we represent the power consumption for training as a stochastic signal fluctuating around a dominant frequency, denoted by $f_0$. Fig.~\ref{siggen}(a) illustrates an example of such a stochastic power profile with the associated time duration and power variables.
To model the variability, we express the time durations for the training cycle and the up/down phases per iteration $i$ as
\begin{align}
    T_{i}^\mathrm{tr}&=\textstyle\frac{1}{f_{0}(1+\xi_i)} \label{trcyc}\\
    T_{i}^\mathrm{up}&=r_{i}^\mathrm{tr}T_{i}^\mathrm{tr},\;\\
    T_{i}^\mathrm{down}&=(1-r_{i}^\mathrm{tr})T_{i}^\mathrm{tr}
\end{align}
where $f_0$ is the baseline fluctuation frequency that is perturbed by $\xi_i$ in iteration $i$. Moreover, $r_i^\mathrm{tr}$ represents the duration ratio for the up phase within iteration $i$. To capture bounded variability across workloads, we model the baseline frequency to be uniformly distributed $f_{0} \sim \mathcal{U}(\underline{f}_0, \bar{f}_0)$, which is fixed for each workload depending on, e.g., model architecture and batch sizes, as reported in \cite{patel2024characterizing, jain2024pal}. Following these studies, for each iteration $i$, its duration randomly deviates from the baseline assuming $\xi_i \sim \mathcal{N}(0, \sigma_\xi^2)$. The phase ratio $r_i^\mathrm{tr} \sim \mathcal{U}(\underline{r}^\mathrm{tr}, \bar{r}^\mathrm{tr})$ is specified based on the empirically observed ranges reported in \cite{latif2024empirical, singhania2024methodology}.
%, following from GPU workload profiling studies. 

In addition, the electricity demands of the two phases are defined as:
\begin{align}
    P_i^\mathrm{up}(t)&=\hat{P}^{\mathrm{tr}}(1+\Delta_i^\mathrm{up}+\eta^\mathrm{up}(t)) \label{trup} \\    
    P_i^\mathrm{down}(t)&=\hat{P}^{\mathrm{tr}}(1-\Delta_i^\mathrm{down}+\eta^\mathrm{down}(t)) \label{trdp}
\end{align}
where $\hat{P}^{\mathrm{tr}}$ denotes the nominal up-phase power consumption. Additionally, for iteration $i$, $\Delta_i^\mathrm{up}$ and $\Delta_i^\mathrm{down}$ denote the baseline power deviations, while $\eta^\mathrm{up}(t)$ and $\eta^\mathrm{down}(t)$ the intra-phase stochastic deviations.
Slow variations in baseline demand, caused by GPU state changes and runtime effects, have been shown to follow an approximately symmetric distribution around the mean, namely $f_0$ \cite{patel2024characterizing, latif2024empirical}. Therefore, we assume a Gaussian distribution for $\Delta_i^\mathrm{up} \sim \mathcal{N}(0, \sigma_{\Delta}^2)$ and $\Delta_i^\mathrm{down} \sim \mathcal{N}(\mu_\Delta, \sigma_{\Delta}^2)$, where $\mu_\Delta$ is the nominal power difference between the two phases. There also exist fine-grained power variations at sub-milliseconds within each phase, due to heterogeneous GPU kernels and communication operations %,  symmetric sub-millisecond deviations 
\cite{patel2024characterizing, singhania2025fingrav}.
Consequently, we represent them as Gaussian random processes, with instantaneous distributions such as $\eta^\mathrm{up}(t) \sim \mathcal{N}(0, \sigma_{\eta,\mathrm{up}}^2)$ and $\eta^\mathrm{down}(t) \sim \mathcal{N}(0, \sigma_{\eta,\mathrm{down}}^2)$.

We further discuss the parameter choices for these probability models, which are summarized in Table~\ref{parameterselection}. For the nominal frequency, recent grid disturbance monitoring reports have linked AI datacenter operations to oscillations near 1 Hz \cite{patel2024characterizing, tesla2025, nerc2025char}. Thus, in our tests, we will set the baseline range to be [0.5,~1.5] Hz. The coefficient of variation in GPU resource utilization reported in \cite{sinha2022not} motivates $\sigma_\xi=0.1$. GPU profiling studies indicate that compute-heavy phases typically occupy 55\% to 80\% of each cycle, leading to $\underline{r}_\mathrm{tr}=0.55$ and $\overline{r}_\mathrm{tr}=0.8$ \cite{li2022ai, hu2024characterization}. For power-related parameters, $\sigma_{\Delta}=0.05$ and $\mu_\Delta=0.3$ reflect an average 30\% reduction in demand during the synchronization phases \cite{patel2024characterizing}. Finally, the intra-phase variability parameters are set to $\sigma_{\eta,\mathrm{up}} \in [0.02, 0.05]$ and $\sigma_{\eta,\mathrm{down}} \in [0.01, 0.03]$, under the assumption that the compute-intensive up phase exhibits greater short-timescale variability than the down phase. Although these parameter values are selected based on reported measurements and profiling studies, the proposed modeling framework is not restricted to these choices and can accommodate different parameter ranges to represent diverse AI workload characteristics.

\begin{table}[t]
\centering
\caption{Example of AI workload model parameters}
\label{parameterselection}
\begin{tabular}{c|c|c|c|c}
\hline
 & \multicolumn{2}{c|}{Training} & \multicolumn{2}{c}{Fine-tuning} \\ \hline
\multirow{3}{*}{Duration}                                               
& $\underline{f}_0,\bar{f}_0$   & 0.5, 1.5    
& $\underline{f}_1,\bar{f}_1$   & 0.3, 0.7      \\ \cline{2-5}
& $\sigma_\xi$                  & 0.1         
& $\sigma_\zeta$                & 0.1      \\ \cline{2-5}
& $\underline{r}^\mathrm{tr},\bar{r}^\mathrm{tr}$  & 0.55, 0.8   
& $\underline{r}^\mathrm{ft},\bar{r}^\mathrm{ft}$  & 0.7, 0.9  \\ \hline
\multirow{4}{*}{\begin{tabular}[c]{@{}c@{}}Power\\ demand\end{tabular}}
& $\sigma_\Delta$      & 0.05         
& $\sigma_\delta$      & 0.03     \\ \cline{2-5}
& $\mu_\Delta$         & 0.3          
& $\mu_\delta$         & 0.8     \\ \cline{2-5}
& $\sigma_{\eta,\mathrm{up}}$    & [0.02,~0.05] 
& $\sigma_{\eta,\mathrm{tail}}$  & [0.01,~0.03]     \\ \cline{2-5}
& $\sigma_{\eta,\mathrm{down}}$  & [0.01,~0.03] 
& $\sigma_{\eta,\mathrm{idle}}$  & [0.005,~0.02]     \\ \hline
\end{tabular}
\end{table}

\subsection{Fine-tuning Workload Modeling}

Fine-tuning workloads also exhibit phase-based periodic power consumption patterns, but their characteristics differ from training workloads. The fine-tuning process typically consists of initialization, early training, late training, and shutdown phases \cite{li2024unseen}. Among these, only early and late training phases contribute meaningfully to overall power consumption. In the early training stage, alternating high-power \textit{tail phases} and low-power \textit{idle phases} form a periodic profile similar to training workloads. In contrast, the late training stage exhibits sustained high power consumption with minimal fluctuations, due to reduced learning rates. Given the limited durations of the other phases and the reduced variability in the late training, this study focuses on modeling the early training stage, which dominates the demand for fine-tuning workloads.

The early training process can be represented with the same underlying structure as large-scale training, as shown in Fig.~\ref{siggen}(b). The cycle structure of the $i$-th iteration is defined as:
\begin{align}
    T_i^\mathrm{ft} &= \textstyle \frac{1}{f_{1}(1+\zeta_i)}, \\
    T_i^\mathrm{tail} &= r_i^\mathrm{ft} T_i^\mathrm{ft}, \;\\
    T_i^\mathrm{idle} &= (1 - r_i^\mathrm{ft}) T_i^\mathrm{ft}
\end{align}
where $f_1$ is the baseline fluctuation frequency with an iteration-based deviation of $\zeta_i$, and $r_i^\mathrm{ft}$ is the ratio of the tail phase. Similarly to modeling large-scale training, empirical studies suggest the frequency variables to follow $f_{1} \sim \mathcal{U}(\underline{f}_1, \bar{f}_1)$ and $\zeta_i \sim \mathcal{N}(0, \sigma_\zeta^2)$. %accounts for symmetric fluctuations around $f_1$. 
The phase ratio $r_i^\mathrm{ft} \sim \mathcal{U}(\underline{r}^\mathrm{ft}, \bar{r}^\mathrm{ft})$ is also used to capture the variation in the compute-to-communication ratio.
Power consumption during tail and idle phases can be defined similarly as:
\begin{align}
    P_i^{\mathrm{tail}}(t) &= \hat{P}^{\mathrm{ft}}(1 + \delta_i^{\mathrm{tail}} + \eta^{\mathrm{tail}}(t)) \\
    P_i^{\mathrm{idle}}(t) &= \hat{P}^{\mathrm{ft}}(1 - \delta_i^{\mathrm{idle}} + \eta^{\mathrm{idle}}(t)).
\end{align}
Here, $\hat{P}^{\mathrm{ft}}$ is the nominal demand for the tail-phase with similar power deviation terms as in  \eqref{trup}-\eqref{trdp}. In addition, probabilistic modeling for the deviation terms also follows with 
$\delta_i^\mathrm{tail}\sim\mathcal{N}(0, \sigma_{\delta}^2)$ and $\delta_i^\mathrm{idle}\sim\mathcal{N}(\mu_\delta, \sigma_\delta^2)$ for slow variations. Fast intra-phase variations are modeled by Gaussian random processes as $\eta^\mathrm{tail}(t)\sim\mathcal{N}(0, \sigma_{\eta,\mathrm{tail}}^2)$ and $\eta^\mathrm{idle}(t)\sim\mathcal{N}(0, \sigma_{\eta,\mathrm{idle}}^2)$.
These components are attributed to the same reasons as those studied for large-scale training.

The parameter choices for the fine-tuning workload model are also shown in Table~\ref{parameterselection}. The dominant fluctuation frequency is set to $f_1 \sim \mathcal{U}(0.3, 0.7)$ based on reported measurements of server-level behavior during fine-tuning workloads, where shorter iteration cycles yield higher intrinsic frequencies, but multi-GPU aggregation suppresses high-frequency components \cite{patel2024characterizing}. Inter-iteration duration variability is modeled as $\zeta_{i} \sim \mathcal{N}(0, 0.1^2)$, following the assumption that variability magnitude is comparable to the training workloads. The compute phase fraction is selected as $\underline{r}_\mathrm{ft} = 0.7$ and $\bar{r}_\mathrm{ft} = 0.9$, reflecting the consistently higher proportion of computation relative to communication due to the reduced number of GPUs \cite{li2024unseen}. The nominal tail-phase demand, $P_{\mathrm{ft},0}$, is chosen to be lower than the training nominal $P_{\mathrm{tr},0}$ to account for smaller datasets and reduced batch sizes. Intra-phase fluctuations are specified as $\sigma_{\eta,\mathrm{tail}} \in [0.01, 0.03]$ and $\sigma_{\eta,\mathrm{idle}} \in [0.005, 0.02]$. At the same time, iteration-level baseline shifts are set to $\sigma_\delta=0.03$ and $\mu_\delta=0.8$ to reflect shorter and less frequent communication phases with consistently low power consumption. Similar to training workloads, these parameters can be adjusted to represent different workload configurations and hardware settings.

\subsection{Aggregated AI Workloads for Datacenter Power Profiling}
To perform power system dynamic studies, we need to model the total datacenter load to explicitly capture stochastic power fluctuations from AI workloads. Since non-server and non-AI loads have relatively stable power consumption profiles \cite{radovanovic2021power}, these components are represented as quasi-constant loads throughout the interval of oscillation studies. Thus, this work attributes the time-varying component of datacenter load solely to AI computational workloads, focusing on high-penetration scenarios in next-generation facilities without active power-fluctuation management.

\begin{figure}[t]
    \centering
    \includegraphics[width=0.9\columnwidth]{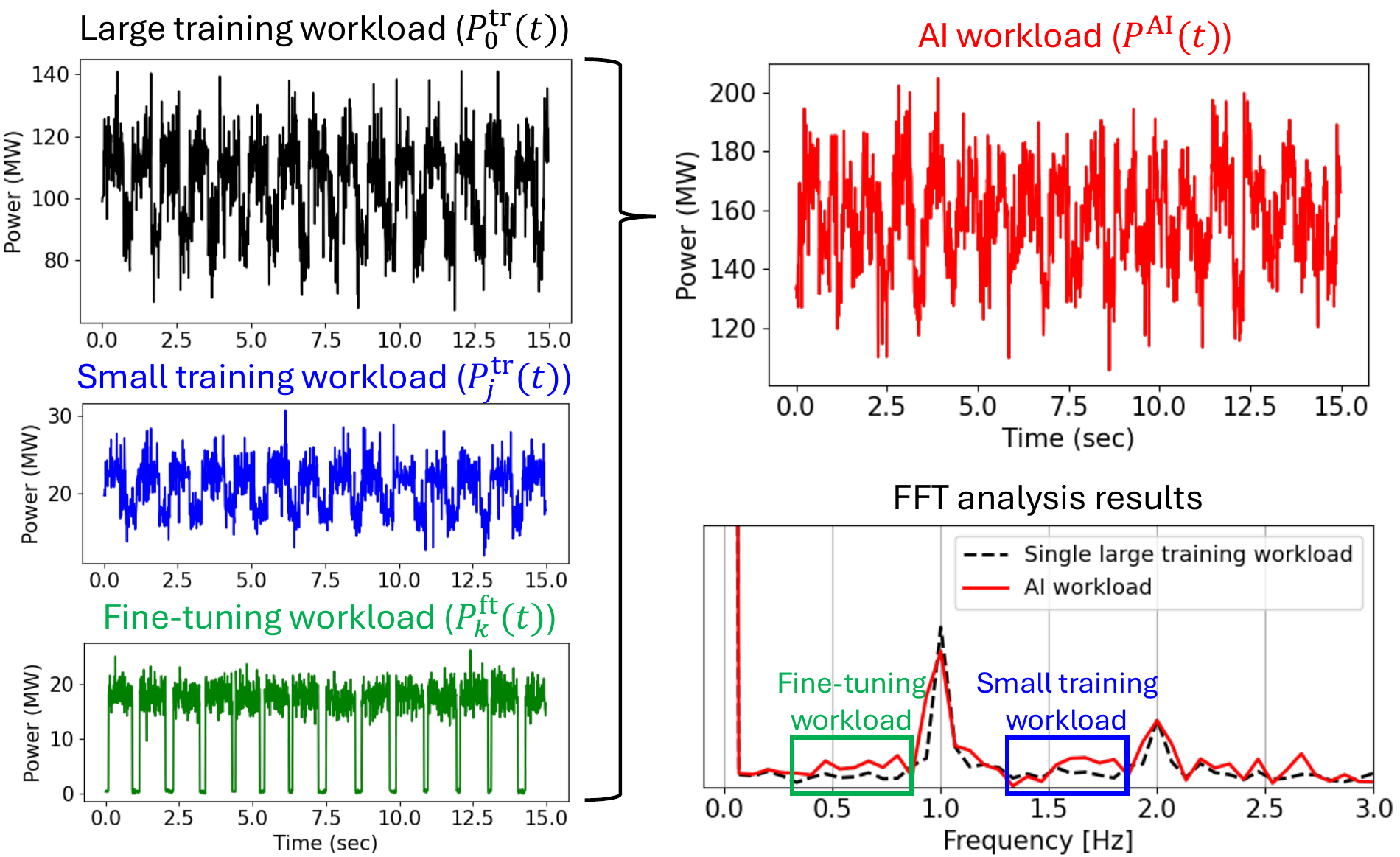}
    \caption{Example of AI workload modeling process.}
    \label{siggen_ex}
\end{figure}

As discussed in Section~\ref{sec2}, these datacenters are typically dominated by a single large-scale training workload, supplemented by smaller-scale training and fine-tuning workloads. Each workload is assumed to operate independently, and the total datacenter load results from the superposition of these stochastic power profiles. Hence, we express the total fluctuating load demand from the datacenter by
\begin{equation}
P^\mathrm{AI}(t) = P^\mathrm{tr}_0(t) + \sum_{j=1}^{N^\mathrm{tr}} P^\mathrm{tr}_j(t) + \sum_{k=1}^{N^\mathrm{ft}} P^\mathrm{ft}_k(t)
\end{equation}
where $P^\mathrm{tr}_0(t)$ denotes the power profile of the dominant and largest training workload. In addition, $P^\mathrm{tr}_j(t)$ corresponds to the $j$-th, small-scale training workload, with a total $N^\mathrm{tr}$ of them; and similarly for $P^\mathrm{ft}_k(t)$ and 
$N^\mathrm{ft}$ representing the fine-tuning workloads. Note that the subscript here stands for the workload index, which is different from the cycle iteration $i$ in \eqref{trup}-\eqref{trdp}. Each workload's power profile is represented by an iteration-based function of time per the aforementioned periodic structure. For example, $P^\mathrm{tr}_0(t)$ can be expressed using \eqref{trcyc}-\eqref{trdp} as:
\begin{equation}
    P^\mathrm{tr}_0(t)=
    \begin{cases}
        P_i^\mathrm{up}(t), & t \in [t_i,\ t_i + T_i^{\mathrm{up}}) \\
        P_i^\mathrm{down}(t), & t \in [t_i + T_i^{\mathrm{up}},\ t_i + T_i^{\mathrm{tr}})
    \end{cases}
    \label{Pexamp}
\end{equation}
where $t_i$ is the start time of the $i$-th iteration. The other two types of workloads follow the same structure, yet at a much smaller nominal power level. Note that transitions between different phases are assumed to occur instantaneously, as they typically occur at a very fast timescale \cite{li2025ai}.
%{\hao this prompts me to think: do we need to consider random $N^{tr}$?}

The relative power contribution of different types of workloads is determined by their nominal electricity demands, namely $\hat{P}^\mathrm{tr}$ and $\hat{P}^\mathrm{ft}$. Based on general AI datacenter resource allocation patterns and future trends \cite{patel2024characterizing, daniel2024AI}, we assume  $\hat{P}_0^\mathrm{tr} : (\sum \hat{P}_j^\mathrm{tr}) : (\sum \hat{P}_k^\mathrm{ft})= 9 : 0.5 : 0.5$ \footnote{This ratio has been chosen based on the reported growth of LLM scale and also discussions with hyperscale datacenter developers. 
Latest reports \cite{joshua2025scaling} suggest large-scale LLM training requires several hundred MWs of power, which will grow with the model size in future. Thus, it is expected emerging GW-level datacenter would allocate the majority of its capacity to large training workloads. Of course, as rapid advancements continue in both LLM architectures and GPU hardware, such proportions may evolve over time. To account for this variability, we further examine the sensitivity of our results to changes in this assumed ratio in Section~\ref{sec4d}.}.
These values have not been directly measured but represent an educated estimate based on the dominance trend of frontier-scale training workloads, such as ChatGPT version updates. Actual values may vary across datacenters, depending on their architectures and operational strategies, and can be readily updated for future studies.
Fig.~\ref{siggen_ex} shows an example of the aggregated AI workload power profile based on our models and assumptions. The large power fluctuation patterns at the second level can be clearly observed, which serve as a forcing source for grid-level oscillation studies.
\begin{remark}[Model stochasticity]
    The proposed model represents a substation-level workload profile corresponding to an entire hyperscale datacenter. In typical LLM training and fine-tuning, the majority GPUs operate in a largely synchronized manner, implying that the aggregated load can be interpreted as a scaled-up version of individual GPU workloads. However, random variability inevitably arises across devices due to differences such as operating conditions and thermal states, as captured by the stochastic elements embedded in our modeling framework. Although the underlying power-electronic mechanisms within a datacenter are not explicitly modeled, the resulting aggregate power profile remains sufficiently representative for assessing system-level interactions with the grid. Hence, we believe our stochastic modeling captures a broad range of workload behaviors, both within a single facility and across heterogeneous datacenters.
\end{remark}
\begin{remark}[Model validation]
     A direct comparison between our model results in Fig.~\ref{siggen_ex} and proprietary real-world measurements is inherently challenging due to lack of data access. Nevertheless, the generated profiles exhibit strong qualitative consistency with publicly available traces reported in the Google Cloud blog~\cite{houle2025balance}, as well as with realistic GPU power measurements in~\cite{li2024unseen, latif2024empirical}. In addition, the temporal features are well aligned with the rack-level power measurements presented in~\cite{choukse2025power}. While dominant fluctuation frequencies differ, they depend on GPU specifications and LLM size which can be adjusted by the parameters in our model. Furthermore, our modeling results have been independently reviewed by researchers from AWS and Meta, who confirmed that the observed behaviors are highly consistent with their practical experience in characterizing real-world datacenter workloads. Collectively, the validity and practical relevance of the proposed modeling framework are well supported.
\end{remark}

\section{Case Studies}\label{sec4}

\begin{table}[t]
\centering
\caption{Baseline Simulation Settings}
\begin{tabular}{c|c|c}
\hline
Simulation Parameters   & WECC            & NPCC          \\ \hline
Base load                         & 60.8 GW         & 27.7 GW       \\ \hline
System inertia ($H_\mathrm{sys}$) & 2.7             & 0.60          \\ \hline
Datacenter penetration            & 11.5 \%         & 10.8 \%       \\ \hline
Number of datacenters             & 7               & 3             \\ \hline
Individual datacenter size        & \multicolumn{2}{c}{1 GW}       \\ \hline
Fluctuation frequency range ($\underline{f}_0$-$\bar{f}_0$)   & \multicolumn{2}{c}{0.5--1.5 Hz} \\ \hline
\{$\hat{P}_0^\mathrm{tr}$ ratio, $N^\mathrm{tr}$, $N^\mathrm{ft}$\}     & \multicolumn{2}{c}{\{90\%, 2, 2\}}        \\ \hline
\end{tabular}
\begin{flushleft}{\footnotesize $\hat{P}_0^\mathrm{tr}$ ratio represents the ratio between nominal demands of large training workload and total AI workload.}\end{flushleft}
\label{baseline}
\end{table}

\subsection{Simulation Settings}
To investigate the impact of AI workloads on power system oscillations, we conduct case studies using the WECC 179-bus system and the NPCC 140-bus system. Both systems are constructed based on the ANDES configuration \cite{cui2020hybrid} with base loads of 60.8 GW and 27.7 GW, respectively. As represented in Table~\ref{baseline}, datacenters are implemented with the penetration level of around 10\% in the baseline case. Each datacenter is assumed to have a nominal capacity of 1 GW, consistent with recent forecasts that a single AI facility could reach this scale \cite{heim2025AI}. Of this capacity, 80\% is modeled as a steady-state component that replaces an equivalent portion of the original aggregated load on the host bus. The remaining 20\% represents the fluctuating AI workload component and is superimposed on the replaced steady-state demand. Accordingly, the datacenter load is represented as a conventional PQ load whose active power varies over time according to the proposed model outputs. The load varies at the simulation timestep of 0.01 s, and the model parameters follow the example values listed in Table~\ref{parameterselection}. This modeling approach preserves the total steady-state load of the original case, ensuring that the base-case power flow solution and the initial operating point for time-domain simulations remain unchanged. The WECC system tests show how system-level and datacenter-level factors govern oscillatory behaviors, while the NPCC system tests verify the persistence of these oscillations and assess mitigation effectiveness in different configurations.

\begin{figure}[t]
    \centering
    \includegraphics[width=0.95\columnwidth]{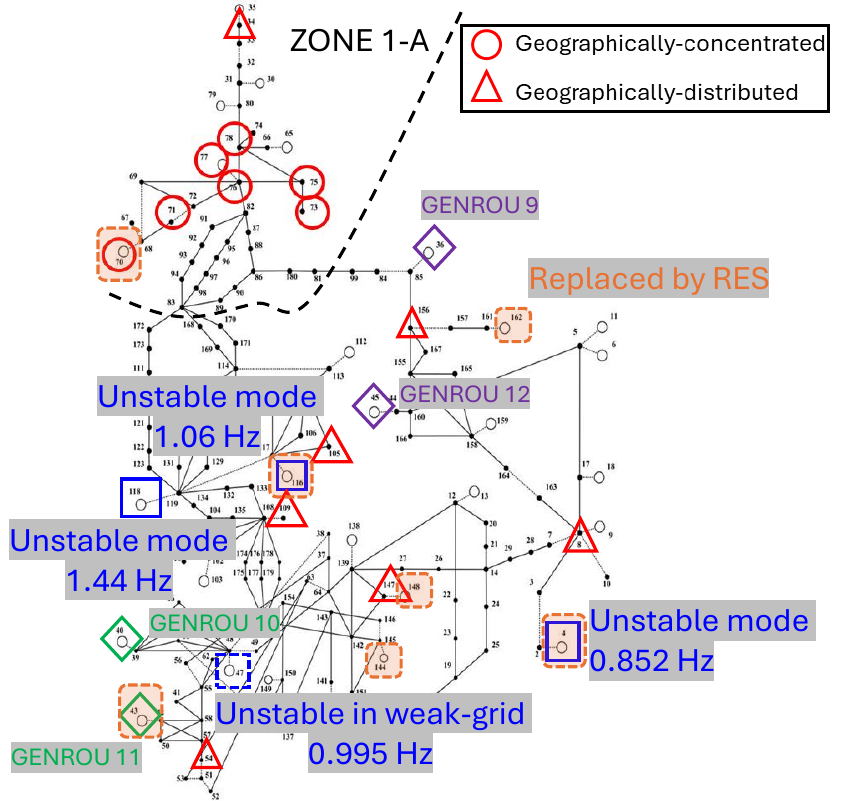} \\ 
    \caption{Topology of the modified WECC 179-bus system with datacenters.}
    \label{top}
\end{figure}

\begin{figure}[t]
    \centering
    \includegraphics[width=0.95\columnwidth]{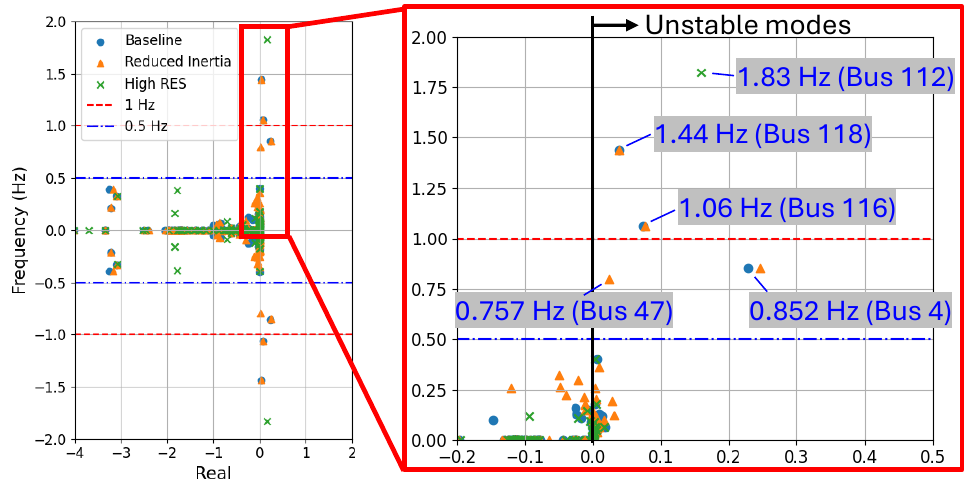} \\
    \caption{Eigen-analysis results of the WECC 179-bus system under different system inertia levels.}
    \label{eig}
\end{figure}
When implementing datacenters, we consider the locations of unstable modes identified through eigen-analysis\footnote{The eigenvalue analysis is performed by linearizing the differential-algebraic equations of the system around the initial operating point. It relies on the initial system configuration, while the Prony analysis results are derived from the measurements obtained from the dynamic simulation. In Fig.~\ref{eig}, a positive real eigenvalue indicates the unstable mode.} using the ANDES tool. In the WECC system, unstable modes at 0.852 Hz, 1.06 Hz, and 1.44 Hz are identified. The locations of these modes are illustrated in Fig.~\ref{top} with the eigen-analysis results in Fig.~\ref{eig}. Accordingly, in the baseline scenario, all datacenters are placed within zone 1-A, which avoids buses with strong participation in these unstable modes, and are randomly assigned within this zone. A distributed deployment scenario is also considered, where datacenters are located on the buses marked by red triangles in Fig.~\ref{top}. In the NPCC system, unstable modes exist mostly below 0.5 Hz, which will be further discussed in Section~\ref{sec4d}. Therefore, only the distributed configuration is examined for this system.

\begin{table}[t]
\centering
\caption{Scenarios for Comparison in the WECC 179-bus System}
\begin{tabular}{cc|l}
\hline
\multicolumn{2}{c|}{Factors}  & \multicolumn{1}{c}{Scenario \& Parameters}  \\ \hline \hline
\multicolumn{3}{c}{System-level factors}  \\ \hline \hline
\multirow{3}{*}{\#1} & \multirow{3}{*}{\begin{tabular}[c]{@{}c@{}}System inertia\\ \& intrinsic mode\end{tabular}} & Baseline ($H_\mathrm{sys}=2.7$) \\ \cline{3-3}
                     &  & Reduced SGs' inertia constant ($H_\mathrm{sys}=1.36$) \\ \cline{3-3}
                     &  & Replaced SGs by RES  ($H_\mathrm{sys}=0.79$) \\ \hline
\multirow{3}{*}{\#2} & \multirow{3}{*}{\begin{tabular}[c]{@{}c@{}}Datacenter\\ penetration\end{tabular}}      & Baseline (11.5\%) \\ \cline{3-3}
                     &   & High penetration level (23\%) \\ \cline{3-3}
                     &   & Extra-high penetration level (34.5\%) \\ \hline \hline
\multicolumn{3}{c}{Datacenter-level factors} \\ \hline \hline
\multirow{2}{*}{\#3} & \multirow{2}{*}{\begin{tabular}[c]{@{}c@{}}Datacenter\\ sizing \& number\end{tabular}}     & Baseline (1 GW$\times$7$=$7 GW) \\  \cline{3-3}
                     &   & Larger size (2.3 GW$\times$3$=$7 GW) \\ \hline
\multirow{3}{*}{\#4} & \multirow{3}{*}{\begin{tabular}[c]{@{}c@{}}Fluctuation\\  frequency\end{tabular}}            & Baseline (0.5--1.5 Hz) \\ \cline{3-3}
                     &             & Narrower frequency range (0.95--1.05 Hz) \\ \cline{3-3}
                     &             & Wider frequency range (0.1--2 Hz) \\ \hline  
\multirow{2}{*}{\#5} & \multirow{2}{*}{\begin{tabular}[c]{@{}c@{}}Geographical\\ location\end{tabular}}             & Centralized (circle in Fig.~\ref{top}) \\ \cline{3-3}
                     &             & Distributed (triangle in Fig.~\ref{top}) \\ \hline
\end{tabular}
\label{simset}
\end{table}

\begin{table}[t]
\centering
\caption{Scenarios for Comparison in the NPCC 140-bus System}
\begin{tabular}{cc|l}
\hline
\multicolumn{2}{c|}{Factors}  & \multicolumn{1}{c}{Scenario \& Parameters}  \\ \hline \hline
\multirow{3}{*}{\#6} & \multirow{3}{*}{\begin{tabular}[c]{@{}c@{}}Mitigation\\ strategy\end{tabular}}  & Without mitigation \\ \cline{3-3}
                     &  & Fluctuation suppression (50--90\%) \\ \cline{3-3}
                     &  & Improved system damping (20--80\%) \\ \hline
\multirow{2}{*}{\#7} & \multirow{2}{*}{\{$\hat{P}_0^\mathrm{tr}$ ratio, $N^\mathrm{tr}$, $N^\mathrm{ft}$\} }      & Baseline ({\{90\%, 2, 2\}}) \\  \cline{3-3}
                     &   & Reduced ({\{60--80\%, 4--8, 4--8\}}) \\ \hline
\end{tabular}
\label{simsetnpcc}
\end{table}

Simulation scenarios are designed to examine how AI-induced oscillations evolve under various factors. In WECC system, system-wide factors, such as system inertia and datacenter penetration level, are evaluated. As the datacenter-level factors, individual datacenter load sizes, load fluctuation frequency, and datacenter geographical placement are compromised. The corresponding scenarios and parameter settings are summarized in Table~\ref{simset}. In the NPCC system, we focus on oscillation-mitigation approaches and workload composition. Specifically, two mitigation approaches are examined, including enhanced systematic damping and fluctuation magnitude suppression, as shown in Table~\ref{simsetnpcc}. Detailed explanations of each scenario are provided in corresponding subsections. All time-domain simulations are performed using the ANDES tool \cite{cui2020hybrid}. Resulting frequency trajectories are analyzed using FFT, Prony analysis, and pseudo energy. By representing the measurement signals as a linear combination of modal components, Prony analysis enables accurate identification of small-signal eigenvalues. As discussed in \cite{almunif2020tutorial}, it generally has a higher reconstruction ability than matrix pencil and eigen-system realization algorithms. The pseudo energy, detailed in \cite{trudnowski2008performance}, serves as an effective metric for characterizing the visibility and impact of specific oscillation modes within the system response.

\subsection{WECC 179-bus System: Impact of System-level Factors}
To assess the impact of datacenter operations on oscillation behavior, we analyze the effects of two key system-level factors: system inertia and datacenter penetration level. The baseline system exhibits a system inertia value of 2.7, and two lower inertial grids are further simulated. The first case proportionally reduces the inertia constants $H$ of all synchronous generators (SGs) by 50\%, yielding a system inertia $H_\mathrm{sys}$ of 1.36 \cite{hartmann2019effects}. The other case replaces 7 SGs by renewable energy sources (RES), resulting in a system inertia of 0.79. The replaced generators are marked in orange in Fig.~\ref{top}, and the RES is modeled using the WECC renewable plant model \cite{force2017central}. As for the datacenter penetration level, we consider a baseline scenario of seven datacenters collectively consuming approximately 7 GW of demand, accounting for 11.5\% of the total system load at 60.8 GW. We further define two higher-penetration scenarios by uniformly increasing the size of each datacenter, keeping the number/location fixed. In the \textit{high penetration} scenario, each datacenter is doubled in size, resulting in a total penetration level of 23\%. In the \textit{extra-high penetration} scenario, datacenter sizes are tripled, corresponding to a penetration level of 34.5\%.

\begin{figure}[t]
    \centering
    \begin{tabular}{cc}
    \includegraphics[width=0.95\columnwidth]{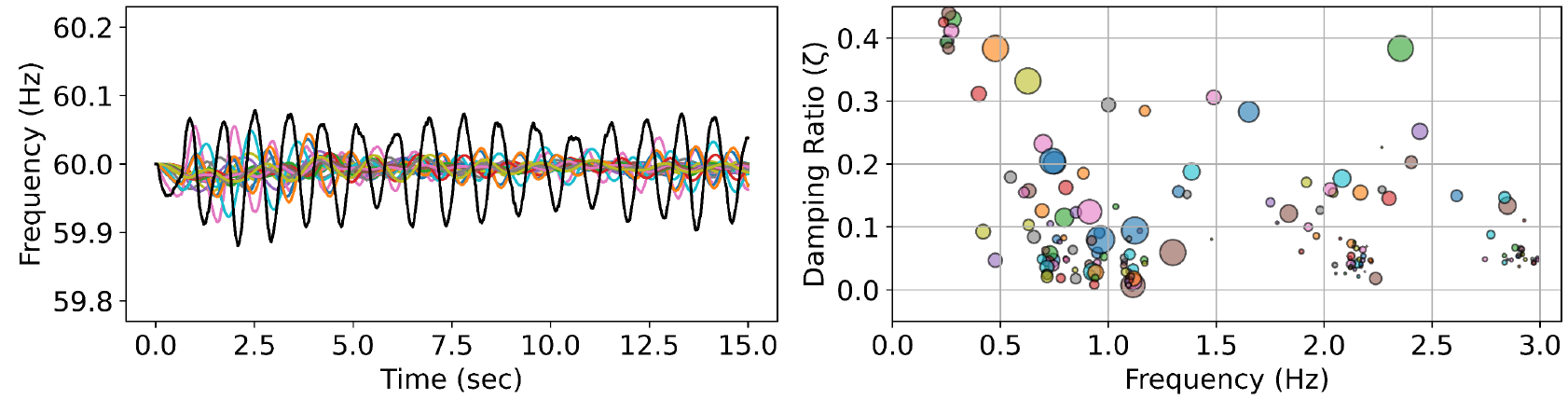} \\
    \small (a) \\
    \includegraphics[width=0.95\columnwidth]{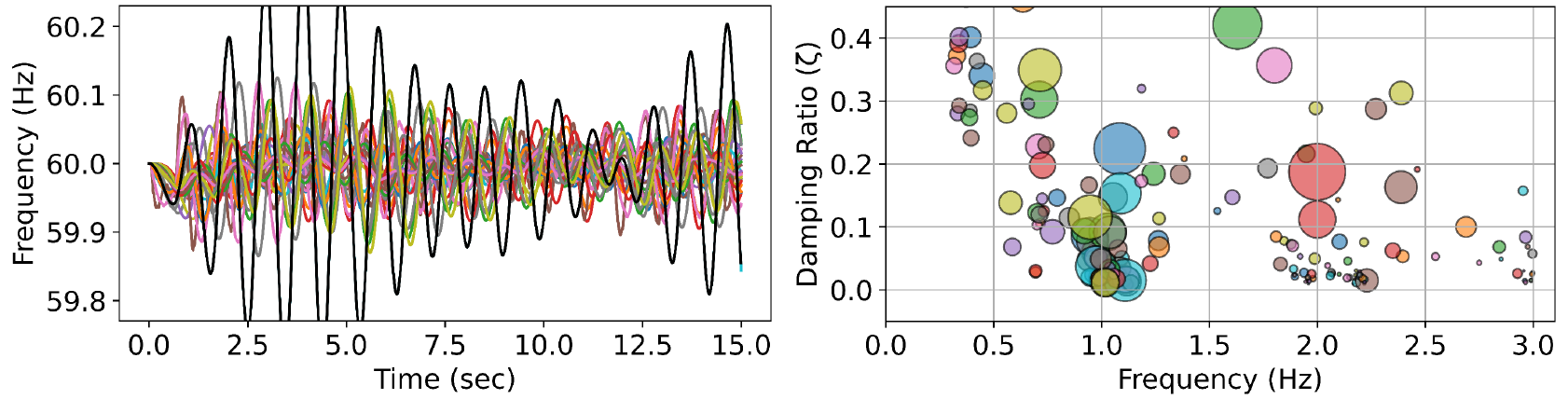} \\
    \small (b) \\
    \includegraphics[width=0.95\columnwidth]{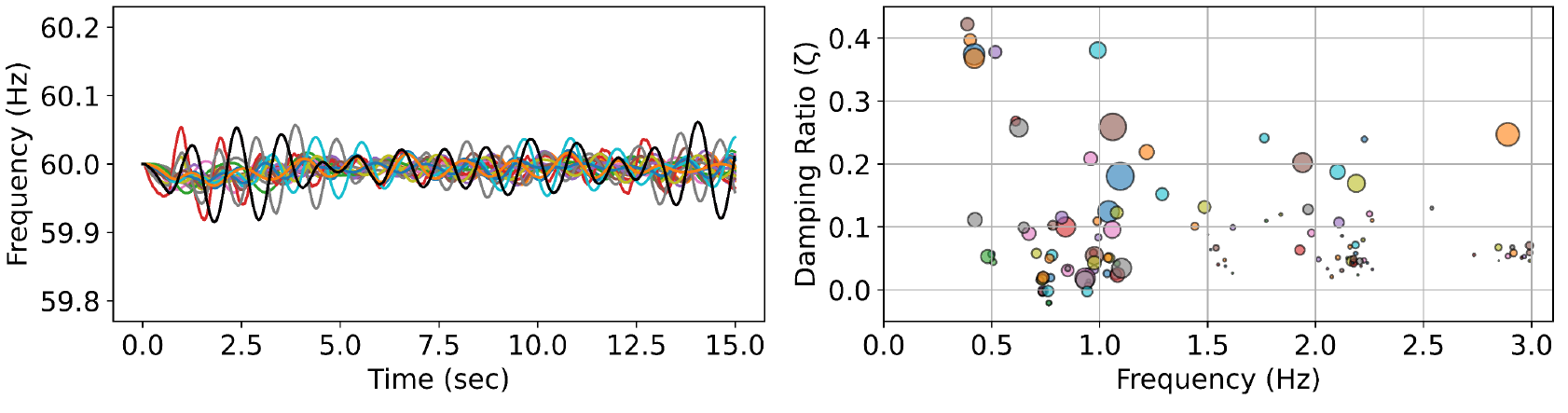}  \\
    \small (c)
    \end{tabular}
    \caption{Frequency trajectory and Prony analysis results according to the system inertia: (a) baseline, (b) reduced generator inertia, and (c) RES replacement cases.}
    \label{gridstrength}
\end{figure}

\noindent \textbf{Factor \#1. System inertia \& intrinsic mode:} Simulation results under different inertia conditions are represented in Fig.~\ref{gridstrength}. The left figures show the generator bus frequencies for each case. The black trace marks the bus with the largest peak-to-peak oscillation. Even in the baseline scenario, sustained oscillation is observed, exhibiting oscillation magnitudes up to approximately 0.2 Hz peak-to-peak. With lower generator inertia in Fig.~\ref{gridstrength}(b), oscillatory behavior intensifies across all generator buses, reaching larger than 0.4 Hz peak-to-peak magnitude. Notably, under the baseline scenario, generator bus frequencies reflect the characteristic shape of the datacenter electricity demand, particularly its higher-frequency components. In contrast, under lower inertia conditions, these features largely disappear from the frequency response. This indicates that as inertia decreases, stronger interactions between datacenter load fluctuations and the power system lead to resonance-like oscillations that dominate system behavior. Surprisingly, although the RES replacement case shows the lowest inertial level, the oscillation magnitudes in Fig.~\ref{gridstrength}(c) are smaller than those of the baseline case. This behavior is explained by the eigen-analysis in Fig.~\ref{eig}, which shows that the RES case lacks an unstable mode around 1 Hz. In contrast, the other cases contain a 1.06 Hz unstable mode associated with Bus 116, located close to the datacenters, enabling resonance with the dominant 1 Hz fluctuation in datacenter demand. The absence of this resonance in the RES case leads to significantly reduced oscillation magnitudes. These results indicate that unstable-mode excitation primarily governs oscillation severity, while lower inertia further aggravates this effect when resonance conditions exist.

Further insights are obtained from Prony analysis results, as shown in the right figures of Fig.~\ref{gridstrength}. The axes represent the modal frequency and damping ratio of oscillatory modes identified from generator bus frequency signals, with the size of each circle indicating its magnitude. In the baseline scenario, oscillation modes are concentrated in the 0.5--1.3 Hz range, with a prominent mode near 1.1 Hz exhibiting the largest amplitude. In the reduced generator inertia case, we can confirm that the oscillation magnitudes are larger than in the baseline case, with similar frequency and damping ratio. This is because the original 1.06 Hz mode, as well as the other unstable modes, shift slightly to the right and become more unstable in the reduced generator inertia case, as shown in Fig.~\ref{eig}. Additionally, a new unstable mode emerges at approximately 0.757 Hz, further degrading system stability. Another notable observation from the eigenvalue analysis is the emergence of numerous unstable modes below 0.5 Hz in the lower inertia case. While these modes exhibit higher damping ratios, they still contribute to overall oscillation amplification, reflected in the enlarged oscillation magnitudes in the Prony analysis results. Finally, the Prony analysis results of the RES case shows reduced modal magnitudes above 0.7 Hz compared to the baseline, consistent with the absence of the unstable mode near 1 Hz. In contrast, modal magnitudes below 0.5 Hz remain comparable to, or even larger than, those of the baseline, reflecting the presence of unstable modes in this low-frequency range. These findings collectively demonstrate that preventing external forcing components that coincide with the system's unstable modes must be a primary consideration for mitigating severe oscillations. In addition, maintaining an adequate system inertia level also plays a key role in keeping oscillation magnitudes within acceptable limits.

\begin{figure}[t]
    \centering
    \includegraphics[width=0.9\columnwidth]{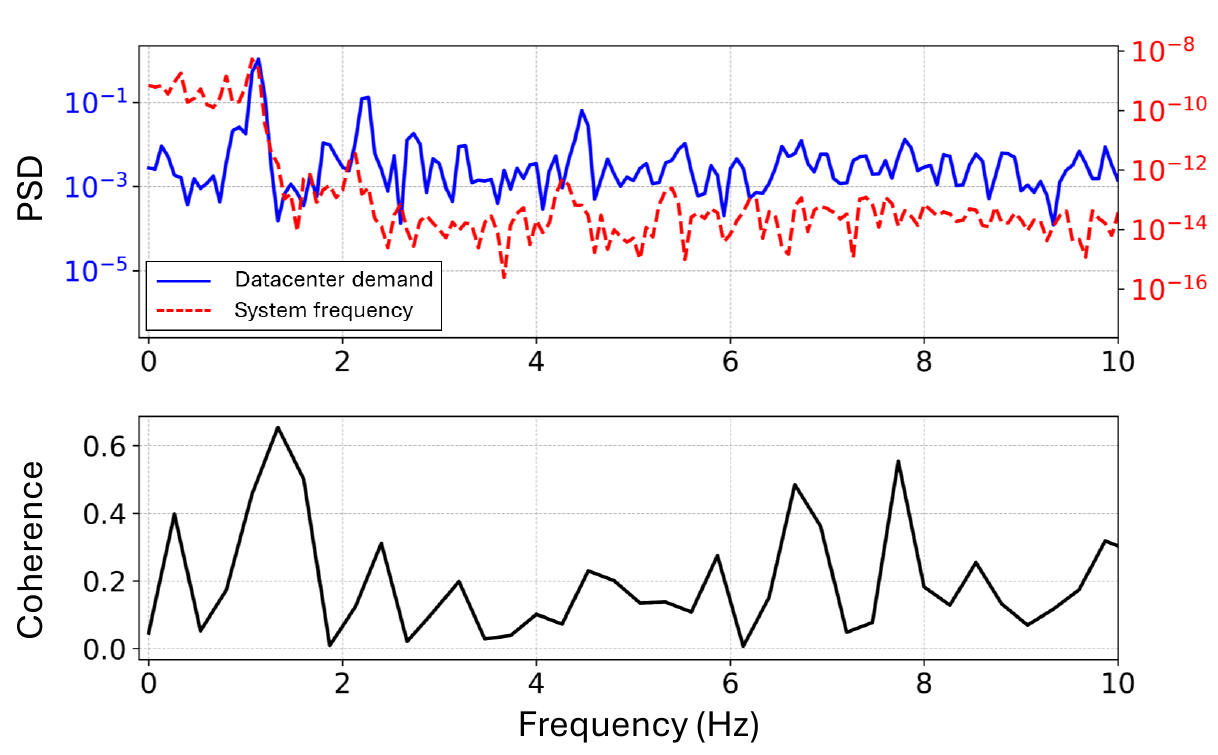} \\
    \vspace{-0.1cm}
    \caption{PSD analysis and coherence of the datacenter demand and system frequency trajectory.}
    \label{psd}
\end{figure}

\begin{figure*}[t]
    \centering
    \begin{tabular}{ccc}
    \multicolumn{3}{c}{\includegraphics[width=1.75\columnwidth]{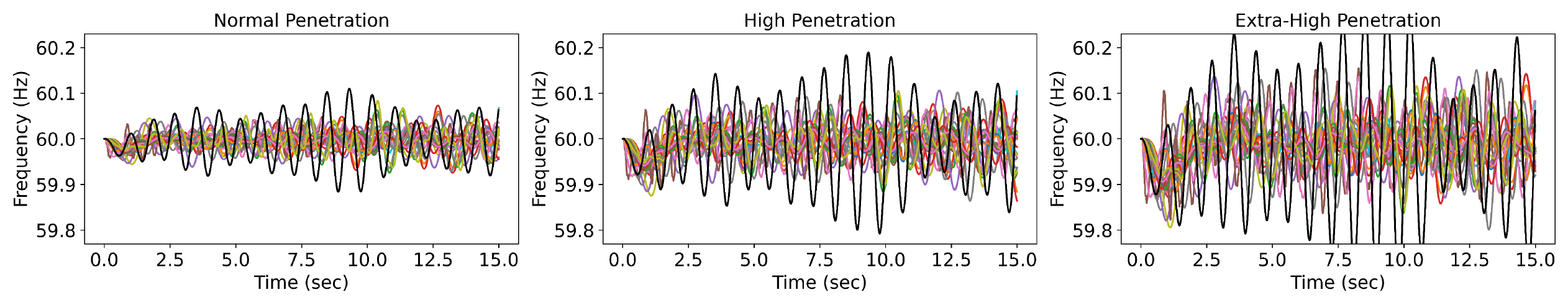}} \\
    \multicolumn{3}{c}{\small (a)} \\
    \includegraphics[width=0.55\columnwidth]{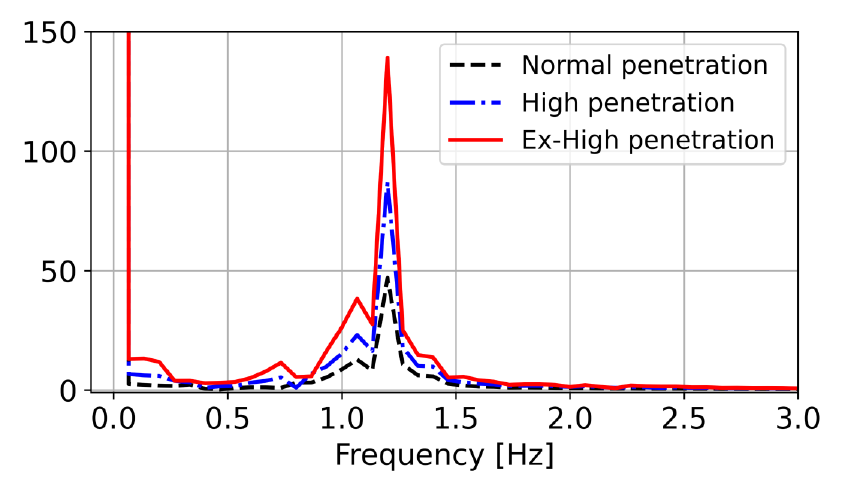} &   \includegraphics[width=0.55\columnwidth]{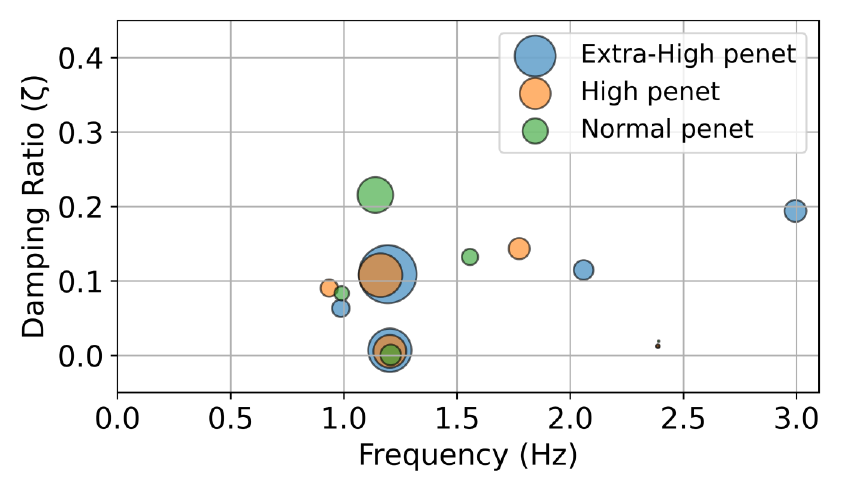} &   \includegraphics[width=0.55\columnwidth]{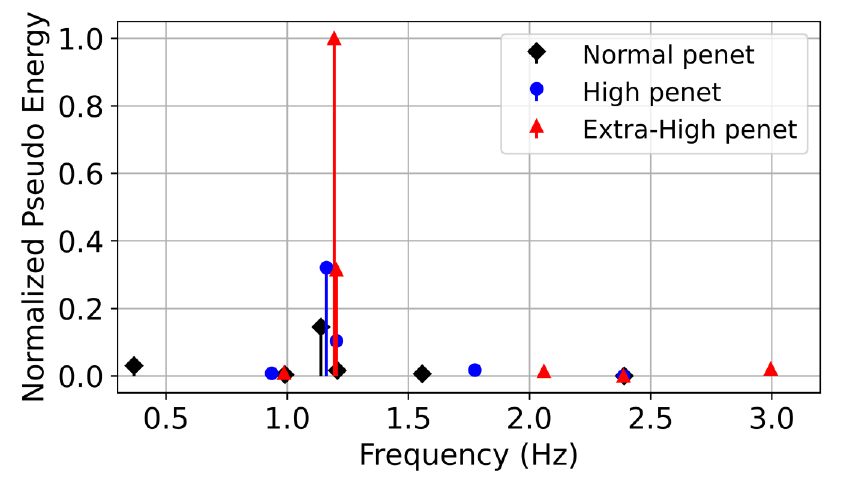} \\
    \small (b)  & \small (c)  & \small (d)   
    \end{tabular}
    \caption{Simulation results according to the datacenter penetration level: (a) electric frequency trajectories, (b) FFT results, (c) Prony analysis results, and (d) pseudo energy.}
    \label{penet}
\end{figure*}

We further perform the power spectral density (PSD) analysis on the datacenter demand and the system frequency trajectory for the baseline case in Fig.~\ref{gridstrength}(a) to distinguish the effects of modal instability and forcing input. As illustrated in Fig.~\ref{psd}, the PSD of the datacenter demand exhibits a dominant peak near 1 Hz with smaller harmonic components, consistent with the FFT results in Fig.~\ref{siggen_ex}. The PSD of the frequency trajectory also shows its largest value around 1 Hz but differs significantly in other frequency ranges. Compared to the datacenter demand, the frequency trajectory shows significantly larger PSD at low frequencies and lower PSD at high frequencies. The elevated PSD value below 1 Hz suggests that weak forcing components are strongly amplified by unstable modes, which is further supported by the low coherence in this region. Meanwhile, the coherence around 1 Hz exceeds 0.6, indicating strong similarity between the forcing signal and the frequency response. This confirms that the 1 Hz oscillation is primarily driven by the datacenter demand and further reinforced by unstable system modes. Above 1 Hz, the low frequency PSD suggests the absence of unstable modes in that region.

\noindent \textbf{Factor \#2. Datacenter penetration:} Simulation results under varying datacenter penetration levels are presented in Fig.~\ref{penet}, where all cases are evaluated under the lower inertia level. As shown in Fig.~\ref{penet}(a), higher penetration consistently amplifies oscillation magnitudes across the system. In particular, under the extra-high penetration, the maximum peak-to-peak oscillation magnitude reaches approximately 0.5 Hz. Among various contributing factors, datacenter penetration level emerges as the primary driver of growth in oscillation magnitude. Unlike the system inertia reduction, higher penetration mainly scales the oscillation amplitude without significantly altering the waveform shape. This distinction is evident in both the FFT results in Fig.~\ref{penet}(b) and the Prony analysis results in Fig.~\ref{penet}(c), obtained from the generator bus exhibiting the most significant oscillation. The dominant oscillation frequency remains approximately 1.2 Hz, and the damping ratio is largely unaffected by penetration level. In contrast, both the FFT amplitude and modal magnitude increase proportionally with penetration level. These trends are further confirmed by the normalized pseudo-energy results, which quantify the mode’s observability in the measurement signal. Results indicate that higher datacenter penetration directly enhances the pseudo energy of the critical mode, reinforcing its dominant influence on overall system oscillatory behavior.

\begin{figure}[t]
    \centering
    \begin{tabular}{c}
    \includegraphics[width=0.85\columnwidth]{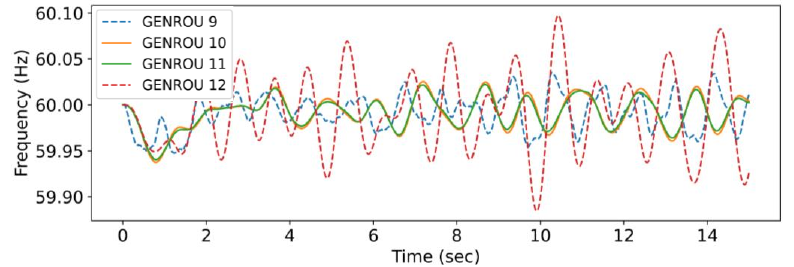} \\
    \small (a) \\
    \includegraphics[width=0.85\columnwidth]{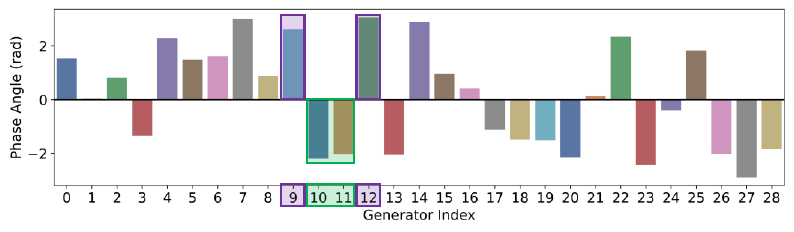} \\
    \small (b)
    \end{tabular}
    \caption{Inter-area oscillation between GENROU 9, 12 and GENROU 10, 11: (a) frequency deviation and (b) mode shape phase angles.}
    \label{interarea}
\end{figure}

Another notable characteristic of datacenter-induced oscillations is that, although the system modes near 1.2 Hz are generally classified as local modes, an inter-area oscillation pattern is observed. As shown in Fig.~\ref{interarea}(a), generators GENROU 9 and GENROU 12 exhibit nearly identical phase responses, while GENROU 10 and GENROU 11 oscillate in anti-phase relative to GENROU 9 and 12. To quantify this behavior, mode shape analysis was performed using Prony analysis applied to the generator frequency deviation signals, focusing on the mode near 1.2 Hz. The phase angles of the resulting mode shape, presented in Fig.~\ref{interarea}(b), show that GENROU 9 and GENROU 12 swing at approximately $+\pi$ radians, whereas GENROU 10 and GENROU 11 swing at approximately $-\pi$ radians. This intense phase separation is a clear indicator of oscillatory separation, consistent with inter-area oscillation behavior. It is further noted that, as depicted in Fig.~\ref{top}(a), GENROU 9 and GENROU 12 are geographically clustered, as are GENROU 10 and GENROU 11, but the two groups are located in distinct network zones. While the frequency response alone does not fully reveal the separation typically associated with classical inter-area modes, the phase-based mode shape analysis confirms the emergence of a zonal inter-area oscillation between these generator groups.

\subsection{WECC 179-bus System: Impact of Datacenter-Level Factors}
We further investigate the impact of key datacenter-level factors on power system oscillations, including datacenter sizing, load fluctuation frequency, and the geographical distribution of datacenters. First, regarding individual datacenter size, the baseline scenario comprises seven datacenters, each rated at 1 GW, collectively consuming a total demand of 7 GW. In contrast, the higher datacenter-size case reconfigures this demand across three larger datacenters, each approximately 3.3 GW, maintaining the same total demand of 7 GW. Second, the frequency of load fluctuation is diversified by adjusting the range of $f_0$ for the dominant training workload. While the baseline case uses a range of 0.5--1.5 Hz, two alternative cases are considered: a narrower range of 0.95--1.05 Hz and a wider range of 0.1--2 Hz. Finally, for the geographical location, we compare a geographically distributed datacenter configuration, as illustrated in Fig.~\ref{top}, against the baseline scenario in which datacenters are concentrated near each other. %Notably, in both cases, the power profile model of each datacenter remains identical to showcase the impact of locational distribution.

\begin{figure}[t]
    \centering
    \begin{tabular}{c}
    \includegraphics[width=0.95\columnwidth]{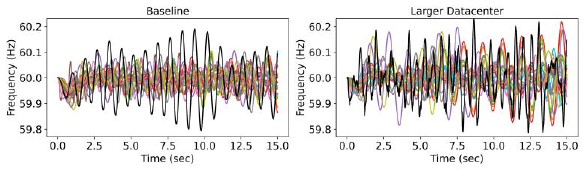} \\
    \small (a) \\
    \includegraphics[width=0.95\columnwidth]{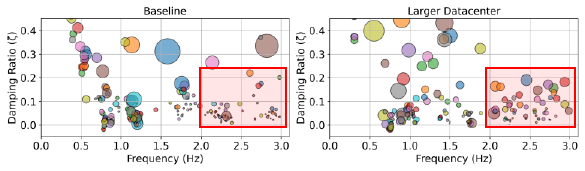} \\
    \small (b)
    \end{tabular}
    \caption{Simulation results according to the individual datacenter size: (a) frequency trajectories and (b) Prony analysis results.}
    \label{indiv}
\end{figure}

\noindent \textbf{Factor \#3. Datacenter sizing:} The experimental results for the datacenter sizing factor are presented in Fig.~\ref{indiv}. Despite the identical total datacenter loading, deploying larger datacenters has resulted in significantly amplified oscillations, as shown in Fig.~\ref{indiv}(a). This amplification effect is particularly evident at buses that do not experience the maximum oscillation in the baseline case. Another important observation is the qualitative change in frequency signals at the bus with the maximum oscillations. While the baseline scenario exhibits conventional oscillatory waveforms, larger datacenter sizing shows more components with frequencies above 1 Hz. This phenomenon indicates that larger individual datacenters exert a stronger influence on the grid, allowing the periodic fluctuations inherent in the datacenter output to manifest more prominently in the system frequency. The Prony analysis results further confirm this behavior in Fig.~\ref{indiv}(b), which reveals an amplification of modes above 2 Hz in the larger datacenter size scenario. However, around the dominant oscillation frequency region near 1 Hz, differences between the baseline and the larger sizing scenario are less pronounced, apart from a moderate increase in mode amplitudes across multiple buses. Although the dominant oscillation shifted from approximately 1.2 Hz in the baseline to 0.8--1 Hz in the larger-size case, this variation is primarily due to stochastic differences in the generated signals.

%\textcolor{magenta}{we need to unify the naming around this paragraph.}

\begin{figure}[t]
    \centering
    \begin{tabular}{cc}
    \multicolumn{2}{c}{\includegraphics[width=0.95\columnwidth]{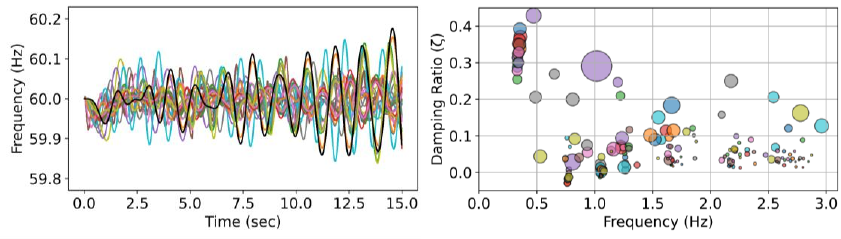}} \\
    \multicolumn{2}{c}{(a)} \\
    \multicolumn{2}{c}{\includegraphics[width=0.95\columnwidth]{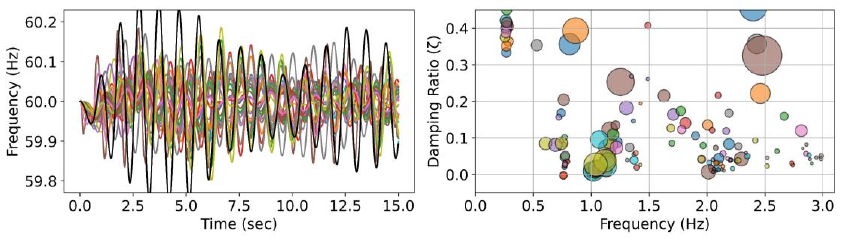}} \\
    \multicolumn{2}{c}{(b)} \\
    \multicolumn{2}{c}{\includegraphics[width=0.95\columnwidth]{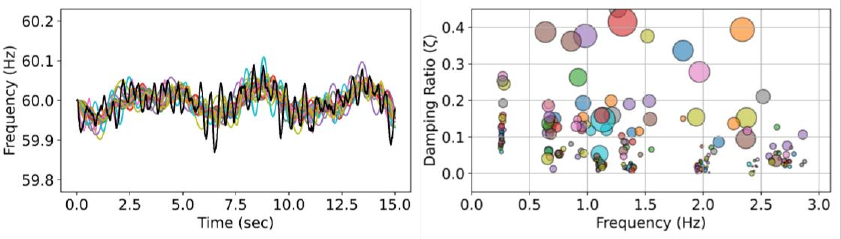}} \\
    \multicolumn{2}{c}{(c)} \\
    \includegraphics[width=0.45\columnwidth]{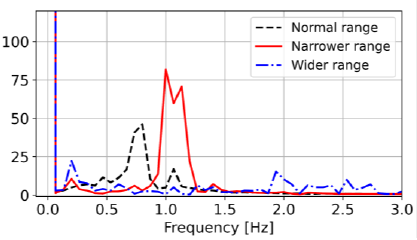} & \includegraphics[width=0.45\columnwidth]{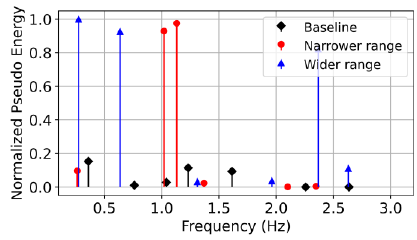}         \\
    (d)      &       (e)
    \end{tabular}
    \caption{Impact of datacenter fluctuation frequency range on system oscillations: frequency trajectories and Prony analysis results under (a) baseline, (b) narrower, and (c) wider ranges. (d) FFT results and (e) pseudo energy.}
    \label{range}
\end{figure}

\noindent \textbf{Factor \#4. Frequency of load fluctuations:} The simulation results under three different fluctuation frequency ranges are presented in Fig.~\ref{range}. Based on the peak-to-peak oscillation magnitude, it is observed that oscillations intensify as the frequency range becomes narrower. This trend can be attributed to the concentration of fluctuations around the 1--1.2 Hz region in the narrower range case, which interacts more strongly with the mode of Bus 116, thereby amplifying the oscillatory response. The Prony analysis results further support this observation. As the frequency range shifts from narrower to baseline, then to wider, the identified modes progressively spread over a broader frequency band, with the mode magnitudes being largest in the narrower-range case. These findings are consistent with the FFT results shown in Fig.~\ref{range}(d), where the narrower range scenario exhibits the highest spectral amplitude near the dominant oscillation frequency. Meanwhile, the wider-range scenario shows a notable distinction from the other cases. Specifically, an additional oscillation component near 0.2 Hz is observed, which is visually apparent in Fig.~\ref{range}(c). Although this oscillation does not appear as a dominant mode in the Prony analysis results or the FFT spectra, it emerges clearly in the pseudo energy distribution shown in Fig.~\ref{range}(e), where it contributes the highest energy content among all frequency bands. This result is partly due to the inherent nature of the pseudo-energy, in which lower-frequency components tend to accumulate larger energy values even when their amplitudes are relatively small. From a power system stability perspective, such wide-area oscillations can be critical. Although their magnitude may appear modest, their prolonged presence can lead to sustained system stress and potential control challenges, making them problematic despite their lower frequency-domain prominence.

\begin{figure}[t]
    \centering
    \begin{tabular}{c}
    \includegraphics[width=0.95\columnwidth]{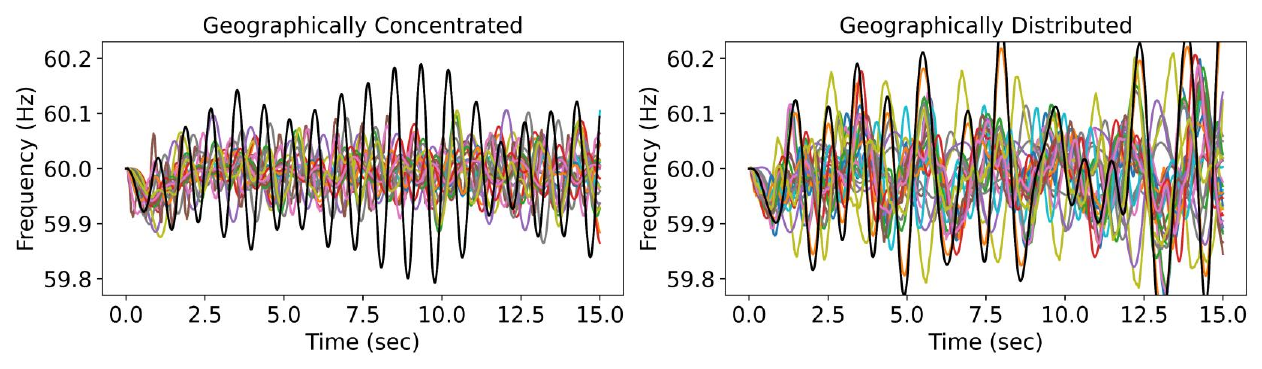} \\
    \small (a) \\
    \includegraphics[width=0.95\columnwidth]{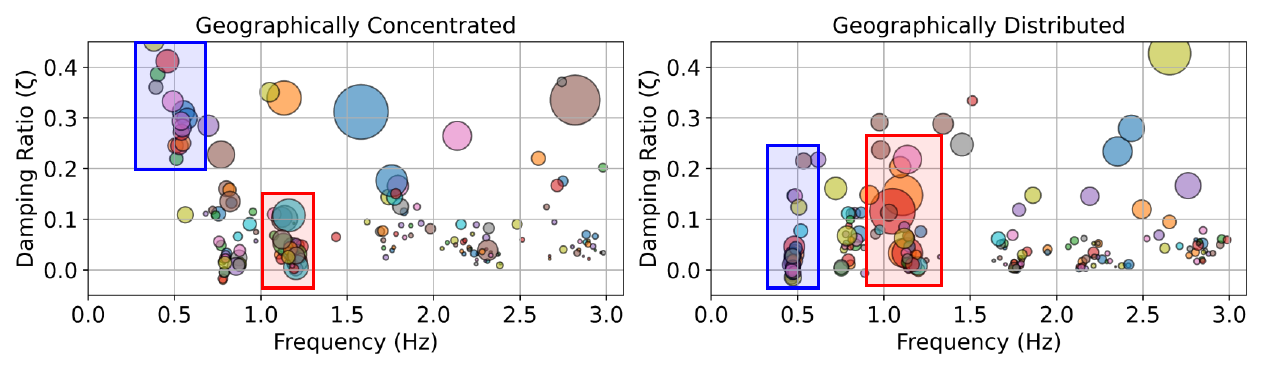} \\
    \small (b)
    \end{tabular}
    \caption{Simulation results according to the geographical location of datacenters: (a) frequency trajectories and (b) Prony analysis results.}
    \label{geo_case}
\end{figure}

\noindent \textbf{Factor \#5. Geographical distribution:} As the final factor, we conduct a case study examining the potential impact of datacenters' geographical distribution. We initially expected that concentrating datacenters in a single area could amplify oscillations due to the aggregated effects of power fluctuations. Surprisingly, as shown in Fig.~\ref{geo_case}, the results revealed the opposite trend. Both the frequency trajectories and the Prony analysis results indicate that, for the geographically distributed datacenter case, oscillations in the 1.2 Hz region become more pronounced than in the concentrated baseline case. An additional noteworthy observation is that the distributed configuration amplifies not only the oscillations around 1.2 Hz but also the oscillations near 0.5 Hz and 0.8 Hz. This can be attributed to interactions between distributed datacenters and multiple local modes within the system. As illustrated in Fig.~\ref{top}(b), the spatially distributed datacenters can interact with the 0.852 Hz mode associated with Bus 4, in addition to the dominant 1.2 Hz mode, likely leading to increased oscillatory responses in both regions. Furthermore, the appearance of a signal component around 0.5 Hz in the Prony analysis can be interpreted as the result of the nonlinear interaction between the two excited modes, generating a difference-frequency oscillation. These findings highlight the complex, aggregated role of datacenter location in shaping multi-mode interactions and amplifying oscillatory levels across multiple frequency bands.

\begin{figure*}[t]
    \centering
    \begin{tabular}{ccc}
    \includegraphics[width=0.635\columnwidth]{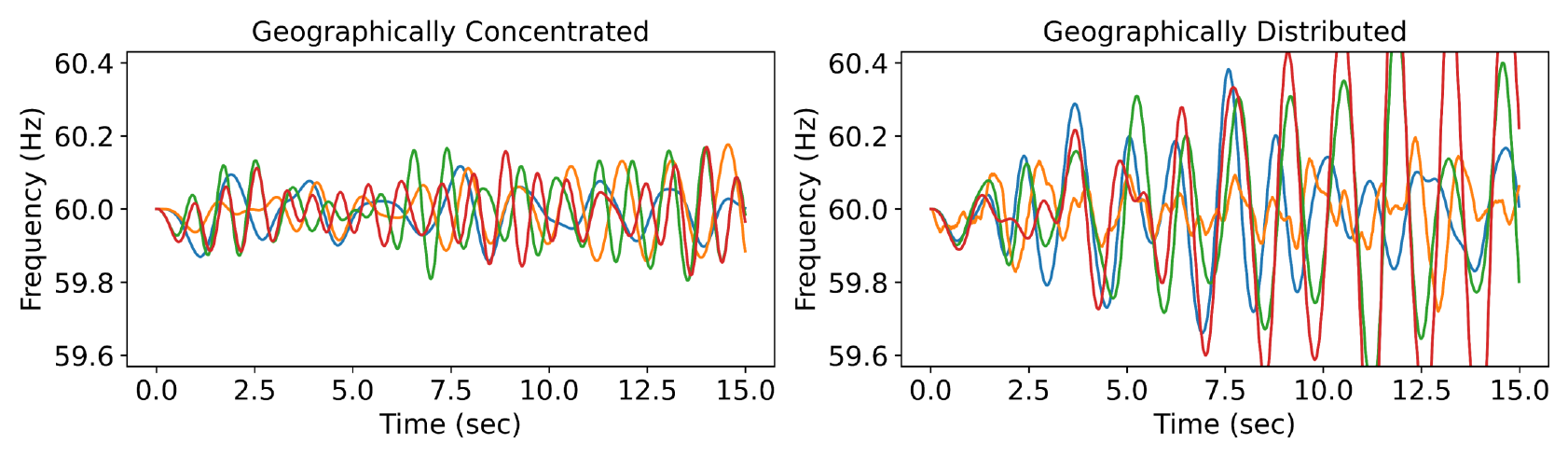} &   \includegraphics[width=0.635\columnwidth]{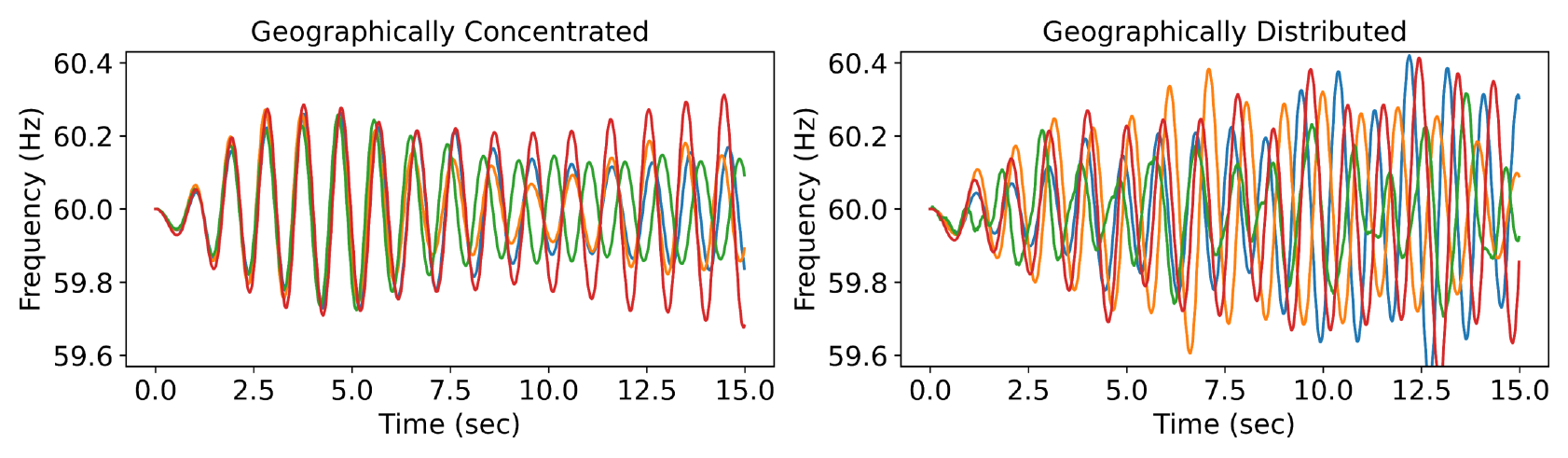} &   \includegraphics[width=0.635\columnwidth]{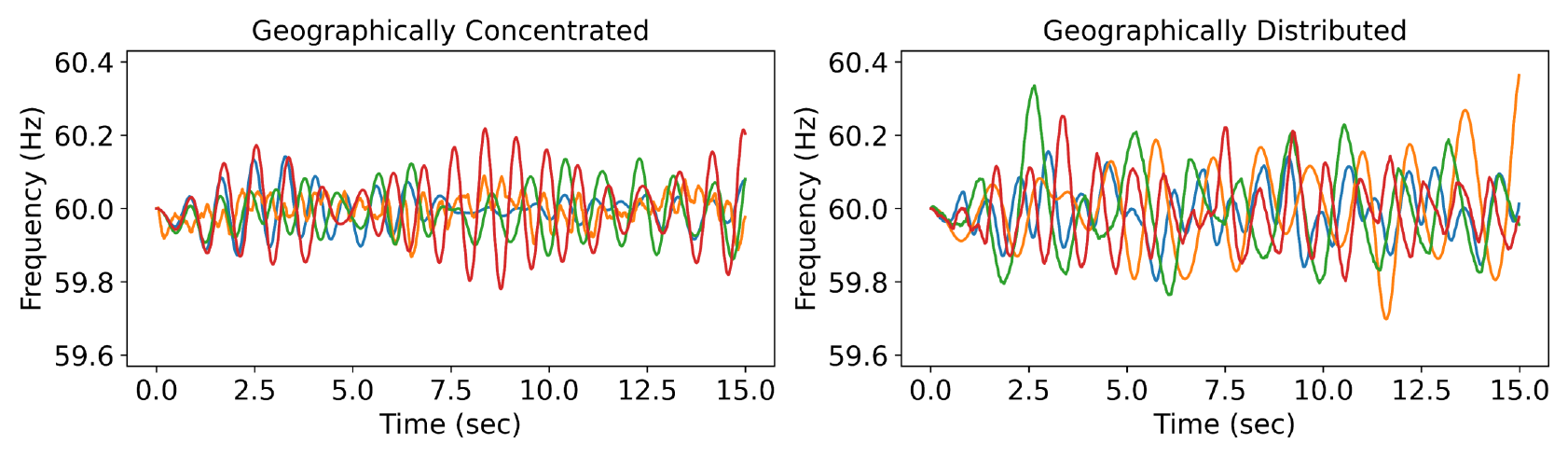}\\
    \includegraphics[width=0.635\columnwidth]{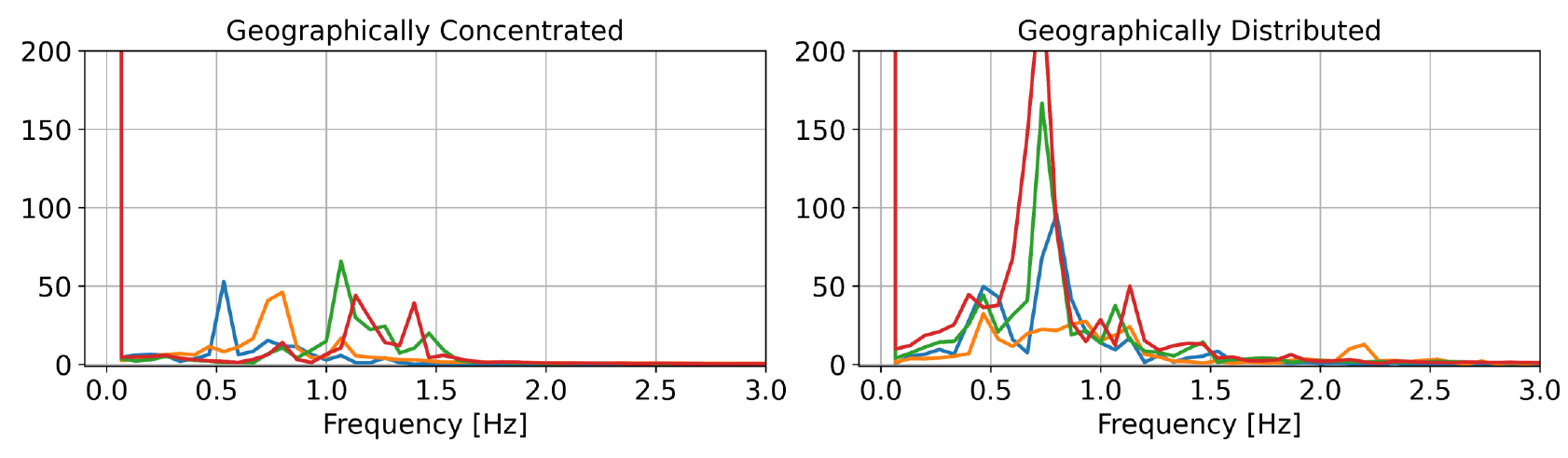} &   \includegraphics[width=0.635\columnwidth]{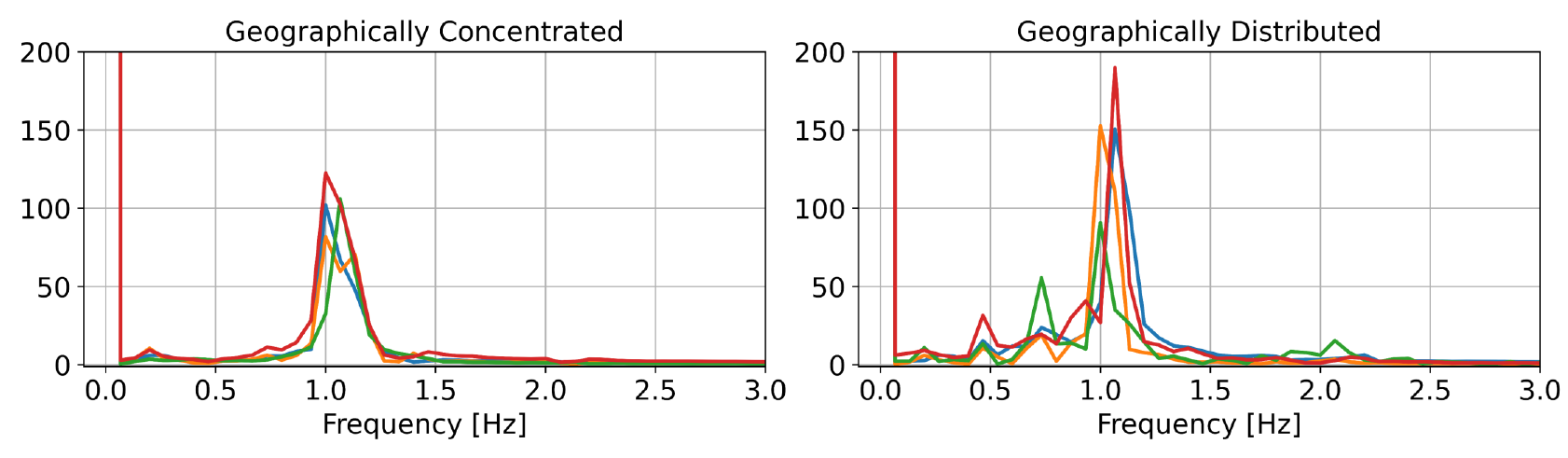} &   \includegraphics[width=0.635\columnwidth]{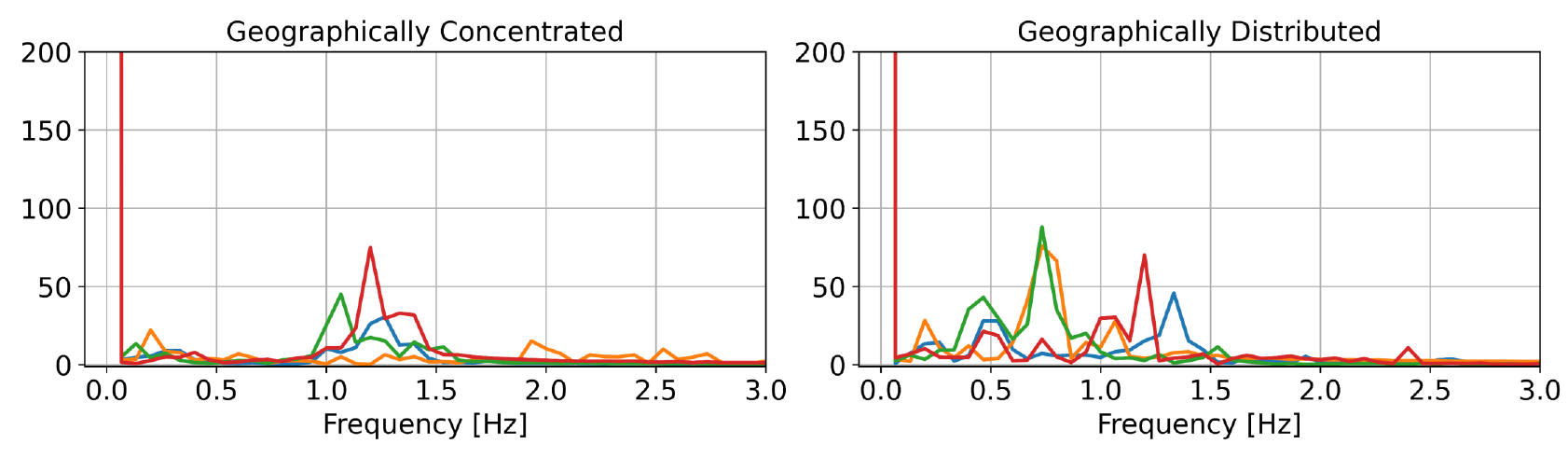}\\
    \small (a)  & \small (b)  & \small (c)   
    \end{tabular}
    \caption{Stochastic variation of frequency trajectories and FFT results for the maximum oscillation bus under (a) baseline, (b) narrower, and (c) wider fluctuation frequency ranges considering different geographical locations.}
    \label{multi_case}
\end{figure*}

To further investigate the potential for diverse mode excitation under geographically distributed datacenter deployments, comparative analyses are conducted across different fluctuation frequency ranges under both concentrated and distributed configurations. Recognizing that stochasticity can result in varying oscillatory outcomes, multiple simulations are performed for each case, and the frequency trajectory and FFT spectrum of the bus exhibiting maximum oscillation amplitude are compared, as illustrated in Fig.~\ref{multi_case}. In the geographically concentrated scenario, consistent with previous discussions, a narrower fluctuation frequency range yields larger oscillation magnitudes, whereas a wider range yields smaller, more dispersed oscillatory responses. In contrast, the distributed deployment scenario exhibits more diverse FFT patterns due to greater stochasticity. Notably, in the narrower-range case, the 0.85 Hz unstable mode remains unexcited. In contrast, in the baseline-range case, interaction with this mode results in larger oscillation magnitudes than in the narrower-range case. In the wider-range scenario, the distributed case exhibits the highest variability across simulation runs, yet consistently shows oscillations in the 0.1--0.3 Hz range, highlighting the sustained presence of low-frequency components under distributed deployments.

\begin{figure}[t]
    \centering
    \includegraphics[width=0.95\columnwidth]{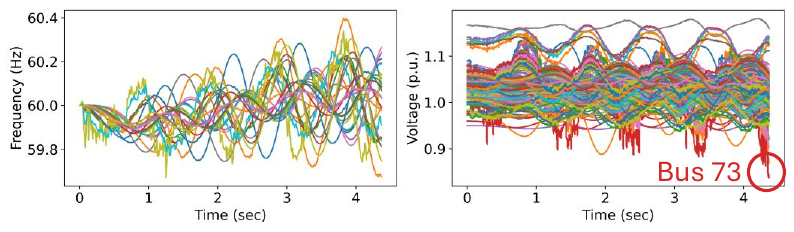} \\
    \caption{Example of simulation failure by the low voltage issue.}
    \label{fail}
\end{figure}

Based on our studies, geographical location is an important decisive factor that affects oscillation behavior. Concentrated datacenter deployments tend to limit the excitation of unstable modes, as their aggregated fluctuations are spatially isolated from critical mode locations. In contrast, distributed deployments increase the likelihood of multiple-mode excitations and broader spectral interactions, particularly under wider fluctuation frequency ranges. Additionally, distributed deployments exhibit stronger stochastic effects, leading to greater variability in oscillation severity and more diverse mode excitation patterns across simulation runs. These spatial coupling and stochastic amplification effects highlight that distributed siting not only exacerbates oscillatory risks but also introduces greater uncertainty in system response. Although concentrated deployments may mitigate oscillatory instability, they raise concerns about voltage stability, as shown in Fig.~\ref{fail}. The observed simulation termination arises from a low-voltage condition that violates the convergence requirements of the ANDES. Once the voltage on Bus 73 falls to 0.837 p.u., the solver encounters an ill-conditioned Jacobian and cannot proceed, resulting in an automatic simulation shutdown. In conclusion, determining the grid-friendly location for AI datacenters requires a holistic approach that considers both their oscillation impact and voltage stability. %Careful planning and analysis are necessary to ensure that large-scale integration of AI datacenters does not compromise overall grid stability, especially as their penetration and workload intensity continue to grow in future power systems.

\subsection{Simulation Results on the NPCC 140-Bus System}\label{sec4d}

\begin{figure}[t]
    \centering
    \begin{tabular}{cc}
    \includegraphics[width=0.47\columnwidth]{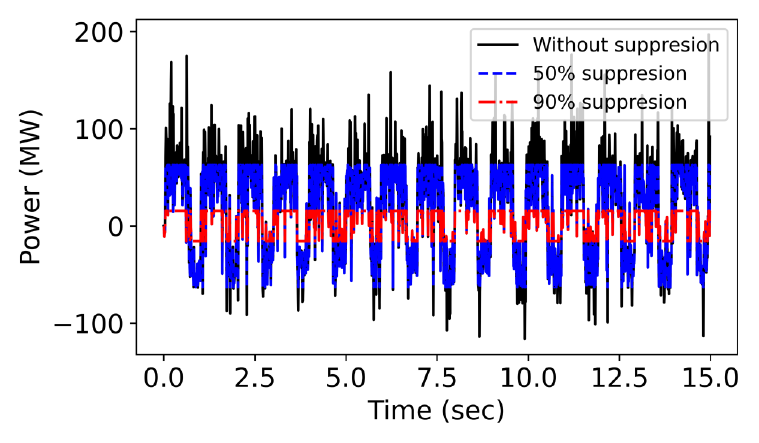} & \includegraphics[width=0.47\columnwidth]{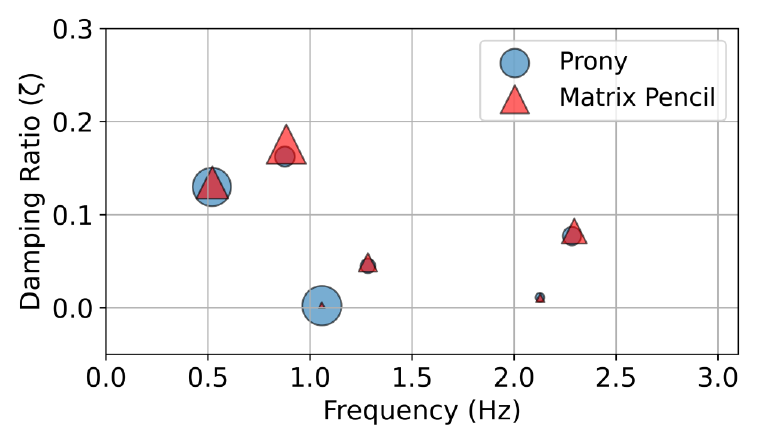}         \\
    (a)      &       (b) \\
    \includegraphics[width=0.47\columnwidth]{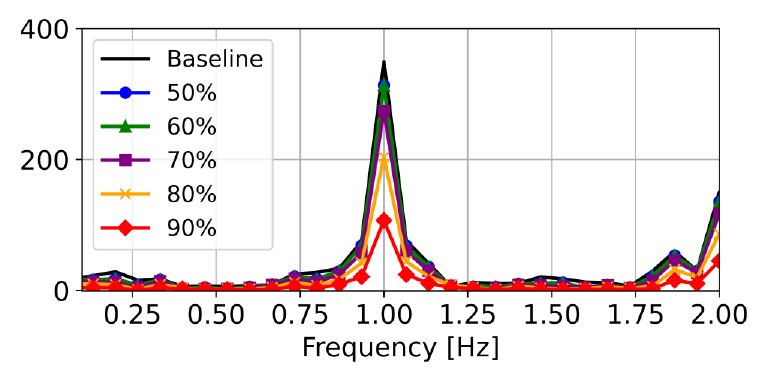} & \includegraphics[width=0.47\columnwidth]{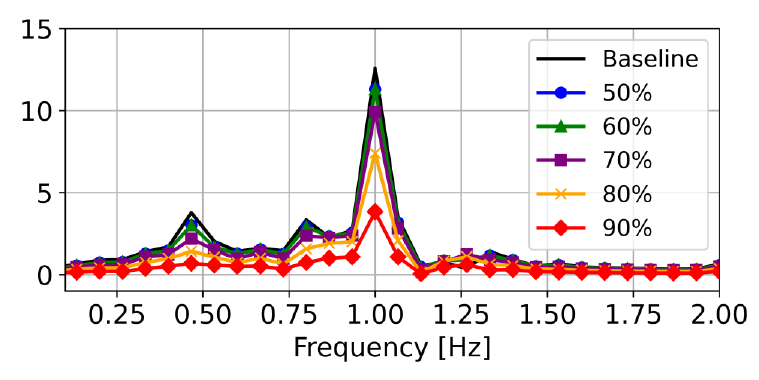}         \\
    (c)      &       (d)
    \end{tabular}
    \caption{Simulation results according to the fluctuation suppression: (a) demand profiles, (b) Prony and matrix pencil analysis results of the baseline case, and FFT results of (c) demand profiles and (d) frequency trajectories.}
    \label{suppression}
\end{figure}

\begin{figure}[t]
    \centering
    \includegraphics[width=0.9\columnwidth]{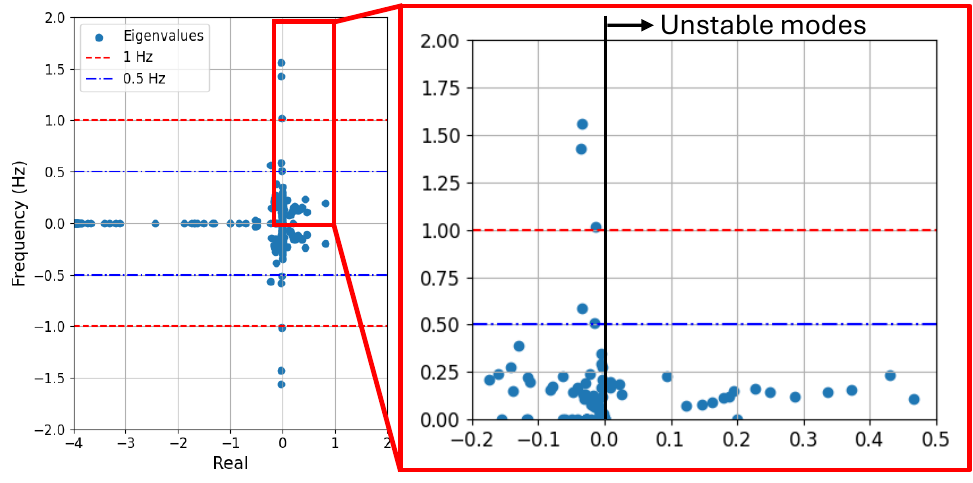} \\
    \caption{Eigen-analysis results of the NPCC 140-bus system.}
    \label{eignpcc}
\end{figure}

To verify the generality of the observed forced oscillations, the baseline scenario is tested on the NPCC 140-bus system, which represents a portion of the eastern interconnection. As shown in Fig.~\ref{suppression}, the dominant oscillation frequency remains near 1 Hz, as also confirmed by the Prony analysis. When comparing the Prony and matrix pencil methods, their extracted frequencies and damping ratios closely match, validating the reliability of the Prony method. The eigen-analysis results in Fig.~\ref{eignpcc} reveal that, unlike the WECC system, the NPCC system contains several unstable modes below 0.25 Hz and none above 0.5 Hz. Consequently, compared to the WECC scenario, the NPCC system exhibits significantly smaller oscillation magnitudes, with the largest peak-to-peak value around 0.1 Hz. These results further demonstrate that forced-oscillation severity is primarily governed by the frequency overlap between the forcing signal and unstable system modes.

\begin{figure}[t]
    \centering
    \begin{tabular}{cc}
    \includegraphics[width=0.47\columnwidth]{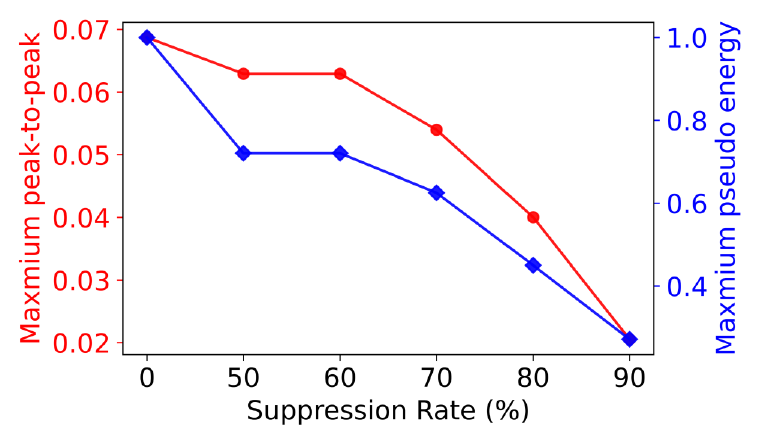} & \includegraphics[width=0.47\columnwidth]{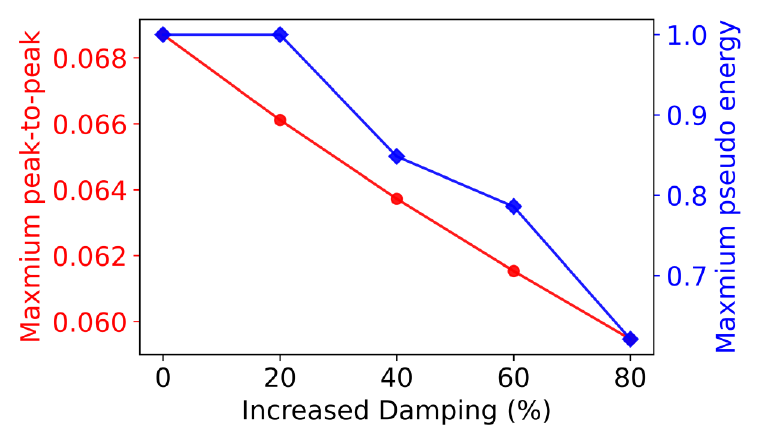}         \\
    (a)      &       (b)
    \end{tabular}
    \caption{Comparison of the maximum peak-to-peak oscillation magnitude and relative pseudo energy according to (a) fluctuation suppression and (b) effective damping.}
    \label{both}
\end{figure}

\noindent \textbf{Factor \#6. Mitigation strategy:} We examine the impact of mitigation approaches to reduce datacenter-induced fluctuations, focusing on two representative strategies: i) smoothing the workload-induced power variations and ii) increasing the system's effective damping. For the former, we assume that the AI workload's maximum fluctuation amplitude is reduced by 50\% to 90\%. The example of suppressed demand profiles is illustrated in Fig.~\ref{suppression}(a) and their FFT results in Fig.~\ref{suppression}(c). 
The suppressed datacenter fluctuation exhibits substantially smaller FFT magnitudes in the 1-Hz band, which naturally reduces the magnitude of the resulting oscillations, as represented in Fig.~\ref{suppression}(d). Moreover, this reduction does not vary linearly with the suppression level. Instead, once exceeding a certain threshold, the effectiveness of mitigation increases marginally. This nonlinear trend is consistently reflected in the maximum peak-to-peak oscillation magnitudes and pseudo-energy values in Fig.~\ref{both}(a). Meanwhile, Fig.~\ref{suppression}(d) shows relatively higher energy content below approximately 0.5 Hz compared than in the demand spectrum. This can be attributed to the unstable low-frequency modes observed in Fig.~\ref{eignpcc}.

To examine the influence of effective damping, we increase the SG damping parameters from 20\% to 80\% and evaluate the resulting system response. As illustrated in Fig.~\ref{both}(b), the oscillation magnitude decreases as the damping increases; however, its mitigation effect is considerably weaker than that achieved through fluctuation suppression. This is mainly because the added damping is distributed among geographically dispersed generators, limiting its localized influence. Although co-locating damping resources near datacenters could enhance effectiveness, the results clearly demonstrate that directly attenuating the forcing signal is a more efficient and robust strategy for oscillation mitigation.

\begin{figure}[t]
    \centering
    \begin{tabular}{cc}
    \includegraphics[width=0.47\columnwidth]{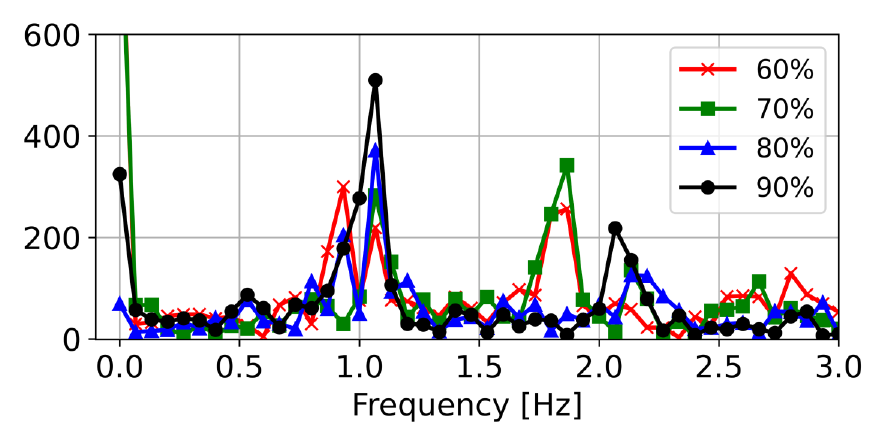} & \includegraphics[width=0.47\columnwidth]{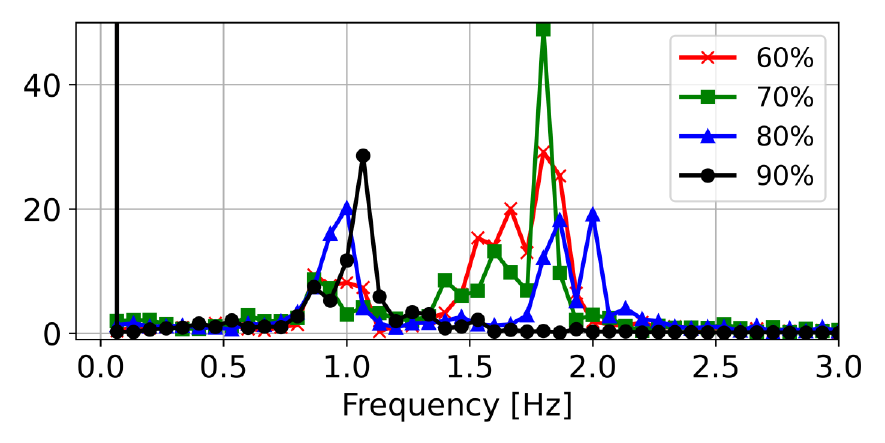}         \\
    (a)      &       (b) \\
    \includegraphics[width=0.47\columnwidth]{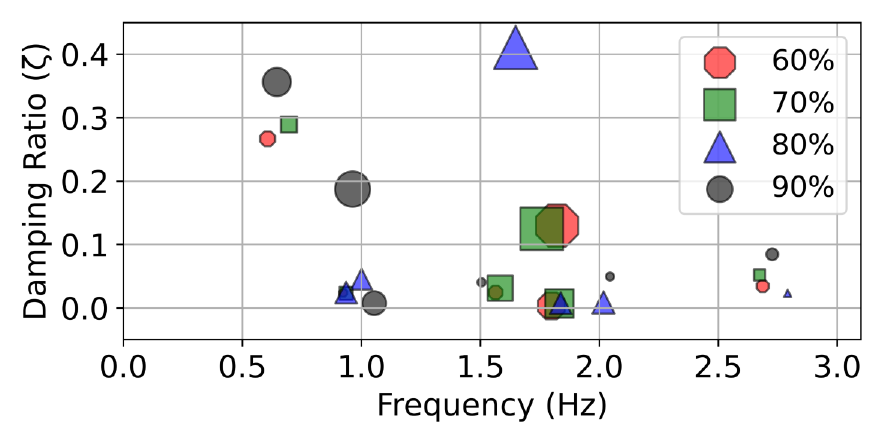} & \includegraphics[width=0.47\columnwidth]{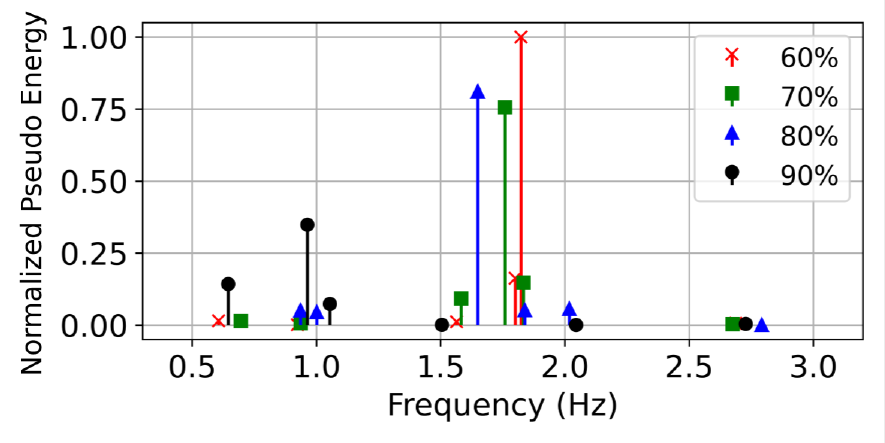}         \\
    (c)      &       (d)
    \end{tabular}
    \caption{Simulation results according to the workload ratio: FFT results of (a) workload and (b) frequency trajectories. (c) Prony analysis results and (d) pseudo energy.}
    \label{tfratio}
\end{figure}

\noindent \textbf{Factor \#7. Workload ratio:} We analyze the impact of large training workload ratio, $\hat{P}_0^\mathrm{tr}$, on oscillatory behavior. This ratio is reduced from the baseline 90\% to 60\%, with the remaining portion reallocated to small training and fine-tuning workloads. For every 10\% reduction in $\hat{P}_0^\mathrm{tr}$, the numbers of these workloads, $N^\mathrm{tr}$ and $N^\mathrm{ft}$, are each increased by 2. As shown in Fig.~\ref{tfratio}(a), decreasing $\hat{P}_0^\mathrm{tr}$ reduces the magnitude around the 1-Hz band, while amplifying components near 0.5 Hz and 1.8 Hz. The amplification around 1.8 Hz can be attributed to the superposition of multiple smaller workloads, which collectively introduce high-frequency components. Frequency trajectory, Prony, and pseudo-energy analyses consistently confirm that the dominant oscillation shifts from the 1-Hz band toward higher-frequency regions as the large-workload proportion decreases. At lower ratios, the 1.8-Hz component exceeds the 1-Hz component in both modal magnitude and energy, indicating constructive accumulation of fine-grained workload variations rather than mutual cancellation.

\begin{figure}[t]
    \centering
    \begin{tabular}{cc}
    \includegraphics[width=0.47\columnwidth]{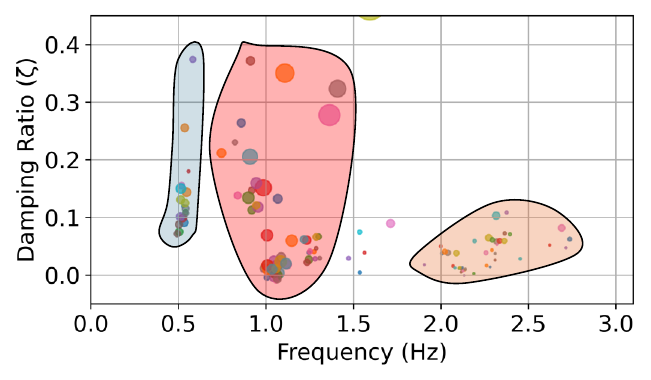} & \includegraphics[width=0.47\columnwidth]{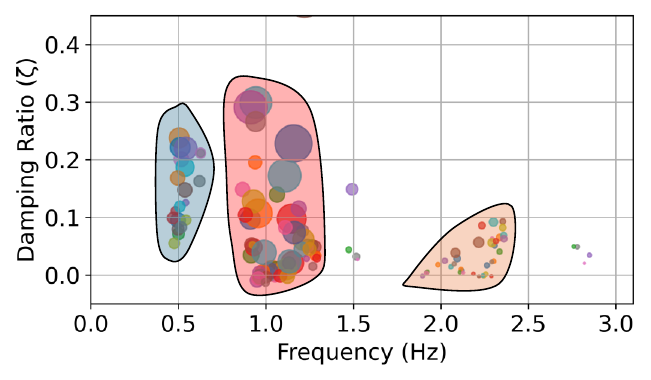}         \\
    (a)      &       (b)
    \end{tabular}
    \caption{Prony analysis results for multiple simulations with (a) normal and (b) narrow frequency ranges.}
    \label{multiplenpcc}
\end{figure}

To further evaluate the impact of stochastic variability in the proposed datacenter model, we run 20 simulations under both normal and narrow workload frequency ranges, as shown in Fig.~\ref{multiplenpcc}. Despite the inherent randomness in individual workload realizations, the resulting system response consistently exhibits a dominant oscillation around 1 Hz, accompanied by smaller oscillatory components near 0.5 Hz and 2 Hz. This persistent pattern indicates that stochastic variations in AI workloads do not diminish the system's susceptibility to datacenter-induced oscillation. The effect becomes even more pronounced in the narrower frequency range case in Fig.~\ref{multiplenpcc}(b), where more coherent forcing leads to stronger and more repeatable oscillatory responses. Notably, these results account not only for randomness within each workload profile but also for inter-workload and inter-datacenter timing differences. The consistent oscillation patterns across all trials suggest that the oscillation risk driven by large-scale AI workloads remains systematically significant even under substantial stochasticity.

\subsection{Oscillation Implications of AI Datacenter Deployment}
Our case studies demonstrate that AI datacenters can act as persistent and intense sources of forced oscillations in power systems when their power fluctuations are not adequately suppressed. Although total datacenter penetration levels are set based on the projection of future deployments, the proportion of AI workloads within these datacenters has been conservatively underestimated. In practical systems, where AI-driven workloads may account for a higher share, the associated risk of oscillation could be significantly higher.

Similar to conventional electromechanical oscillations, the severity of datacenter-induced oscillations is governed by several key system and load characteristics. Lower system inertia, higher datacenter penetration, and larger individual datacenter sizes intensify oscillation severity. Most critically, the impact of these oscillations is dictated by the degree of frequency alignment between datacenter load fluctuations and the system's unstable modes. When the forcing signal resonates with an existing unstable mode, the oscillation magnitudes are dramatically amplified, posing serious risks to system reliability. Geographical siting, fluctuation frequency range, and workload composition influence this alignment either directly or indirectly. From a mitigation perspective, increasing system inertia or damping can alleviate oscillations by limiting their amplitudes once excited. However, our study shows that suppressing load fluctuations at their source is a more effective and robust strategy for preventing forced oscillations.

\section{Conclusion} \label{sec5}
This paper has presented a stochastic power profiling framework that explicitly captures the periodic and high-magnitude fluctuations introduced by large-scale AI workloads in next-generation datacenters. By incorporating workload-specific characteristics, the proposed method models AI-centric datacenters as continuous, frequency-selective forcing sources distinct from conventional ramping loads. Using the WECC 179-bus system and NPCC 140-bus system, we have systematically evaluated forced oscillations induced by datacenters under various factors.
Our results show that severe oscillations can emerge when datacenter-induced fluctuations excite intrinsic unstable modes of the grid. The severity of these oscillations is strongly influenced by the fluctuation’s frequency content and geographical location, as well as by broader system factors such as workload composition and datacenter deployment configuration. Direct suppression of the datacenter load fluctuation is demonstrated to be substantially more effective than indirect system-level measures, highlighting the importance of understanding the characteristics of workload-driven variability when assessing oscillatory risk. These findings emphasize the need to incorporate workload-based electricity demand modeling and forced-oscillation risk assessments into future grid planning and operational studies, particularly as AI-centric datacenters grow in scale and prevalence.
Exciting future research directions open up including the mitigation of power fluctuations at both datacenter- and grid-levels, as well as the extension to modeling a wider time-scale range in the variability of datacenter power consumption. 

\textcolor{magenta}{}

\ifCLASSOPTIONcaptionsoff
  \newpage
\fi
\bibliographystyle{IEEEtran}
\bibliography{bibliography}

\end{document}